\documentstyle  {article}
\begin{document}

\noindent   {\Large{\bf About the nonlocal particle-field matter}}
 
\bigskip\bigskip
\hspace {10 mm} { \parbox {10 cm}
 {{ 
\noindent {\bf I E Bulyzhenkov }

\bigskip \noindent  {\small Institute of Spectroscopy RAS, Troitsk, 142092
Moscow Reg., Russia}
\bigskip
 
\noindent  {\bf Abstract}

\noindent We propose that particles are associated both with localized macroscopic
states at point vertices and with extended microscopic states at all
 vacuum points. The self-fields screen the microscopic particle currents
 everywhere except at the particle vertex, observed as a
point mass/charge carrier of such a nonlocal 
particle-field object. 
All vertices are excluded from the microscopic Maxwell-Lorentz and
Einstein-type equations for elementary objects with continuous particle
 densities. The geodesic particle
motion depends on  a vector force with unified electromagnetic and
gravitational  components. The advanced gravitational wave, which will
 never
come from infinity at finite times, and the retarded electromagnetic
 wave are vector anti-waves. Curved pseudo-Riemannian space-time always
  maintains flat 3D space that is in agreement with the measurements of planetary
   perihelion precession, gravitational light bending, radar echo
    delay, and the nearly
isotropic 2.73K cosmic microwave background. The developed synthesis of
gravity with electrodynamics and the particle with its field corresponds
 to the predicted way of double unification and the Rainich-Misner criterion.

   }}}
 \bigskip \bigskip

\noindent {\bf 1. Introduction}

\bigskip

Pseudo-Riemannian geometry of curved space-time in General Relativity [1] is
determined  by the total system of gravitational masses in the Universe.
Einstein's macroscopic relativity operates with one metric tensor for one
 curved space-time, which is considered as a common manifold for all bodies.
 Three-space is also curved in General Relativity due to Schwarzschild's
  solution for the point mass [2].

However,  the balloon measurements [3] of the nearly isotropic 2.73K cosmic
microwave background radiation strongly favour a spatial flatness of the
Universe. Euclidean three-space may be accepted to modernize relativistic
 gravity, for example  [4],  and this space is quite appropriate for all
 other kinds of interactions. Having employed only spin-1 mediators (photon,
  $W^{\pm}$ and $Z$ mesons, and eight gluons), the Standard Model may
  anticipate the spin-1 graviton in the grand unification.

We consider  different curved four-spaces for different elementary particles
 and their finite ensembles. And we try to justify Newtonian flat space
 and the absolute time rate due to intrinsic metric symmetries of these
 curved four-spaces. We accept that the proper pseudo-Riemannian four-space
  $x^\mu_{_N}$ of one
selected particle N with the mass $m_{_N}$ takes pseudo-Euclidean metrics
(four-interval)
in the absence of all other particles.
Recall that space-time in General Relativity is curved even for the sole 
mass $m_{_N}$ in the Universe.  
Metrics of the curved proper four-space $x^\mu_{_N}$, specified for one
selected mass $m_{_N}$, depends exclusively on a gravitational potential
  of external fields ($B_\mu^{\neq_N}$ in our approach), with $ds^2_{_N}
  (B_\mu^{\neq_N}) = d\tau_{_N}^2(B_\mu^{\neq_N}) - dl^2_{_N} $, $d\tau_{_N}
  (B_\mu^{\neq_N}) = [g^{_N}_{oo}(dx_o + g^{-1}_{_Noo}g^{_N}_{oi}dx^i)]^{1/2}$,
   and $dl_{_N} =
[\gamma^{_N}_{ij} dx^i dx^j]^{1/2}$,  $i,j = 1,2,3$, and $\mu,\nu = 0,1,2,3)$.
 External systems are always different for different masses $m_{_N}$ and
 $m_{_K}$, and metrics $ds_{_N}(B_\mu^{\neq_N})$ and $ds_{_K}(B_\mu^{\neq_K})$
  of the pseudo-Riemannian four-spaces $x^\mu_{_N}$ and $x^\mu_{_K}$,
   respectively, are different in principle if $N\neq K$.

A program goal of this paper is to derive  electrodynamic and gravitational
 equations for one elementary particle N in
its curved four-space  $x_{_N}^\mu$ and to sum these equations over an
ensemble of particles. The world space ${x^i}$ for this ensemble is introduced
 as  a common intersection of the proper 3D-subspaces $x_{_K}^i$ which always
 maintain the universal (Euclidean) geometry within the proper curved
 four-spaces $x_{_K}^\mu$ due to the symmetries $\gamma^{_K}_{ij}
\equiv   g^{_K}_{oi}g^{_K}_{oj}(g^{_K}_{oo})^{-1} -g^{_K}_{ij} = \delta_{ij} $.

At first we shall verify that the accepted tetrad formalism does not contradict
to the intrinsic metric symmetries $\gamma^{_K}_{ij} = \delta_{ij} $. Then we
shall find the pseudo-Riemannian metric tensor $g^{_K}_{\mu \nu }$, which
satisfies to these symmetries under arbitrary gravitational and
electromagnetic  vector potentials and their gauges. This finding
 corresponds in principle to the existing opportunity to use the
 universal three-space
$x^i$ and co-ordinate time $x^o$ as a common manifold $\{x^i; x^o \}$ for
 description of all particles and their fields. It is essential for our
 space and time notions that a 1D absolute
time interval,  $dt_{_K} \equiv 
( \gamma^{_K}_{oo}dx^o dx^o)^{1/2} = |dx^o| = dt >0$, maintains also a
universal form with $\gamma^{_K}_{oo} = \delta_{oo}$ and may be applied
 to  all elementary particles K.

The novel covariant scheme for relativity with flat 3D subspace is
 consistent with all known experiments and observations. In particularly,
 it explains  quantitatively the measured planet perihelion precession,
 the gravitational light bending, and the radar echo delay. Our relativistic
  corrections to Newtonian motion in weak fields coincide with the similar
  corrections of General Relativity, but strong fields in our flat and GR's
   curved three-spaces lead to different dynamics of relativistic matter.

Our "non-curved" three-velocity $v_i\!\equiv\!\gamma_{ij}\!v^j\!=\!\delta_{ij}
\!v^j$ and $v^i \equiv dx^i/d\tau_{_N}$ in flat three-space admits a linear
decomposition of a particle four-momentum, $P_{_N\mu} = m_{_N}V_{\mu}$ into
 an inertial,  $m_{_N}(1-\delta_{ij}v^iv^j)^{-1/2}\{1; -\delta_{ij}v^j\}$,
 and field, $m_{_N}(1-\delta_{ij}v^iv^j)^{-1/2}\{  {\sqrt {g_{oo}} }-1;
 -{\sqrt {g_{oo}} }  g_i  \}$, summands that prohibited in General
 Relativity.
And a linear superposition of gravitational, $ m_{_N}B_{\mu}(x)$,
and electromagnetic, $q_{_N}A_{\mu}(x)$, potentials takes place under flat
 space in the canonical four-momentum of the particle N with the charge
 $q_{_N}$.

By following de Broglie's departure from the point classical particle,
we put into consideration at all vacuum point $x\neq \xi_{_N},\xi_{_K}$
 the extended microscopic states of the particle N, located at one vertex
  point $\xi^\mu_{_N}$ under its averaged, macroscopic state. These extended
microscopic  states may be formally considered as outgoing and incoming
virtual fluctuations which induce locally outgoing and incoming elementary
 fields, $q_{_N}f^{_N+}_{\mu\nu}(x)_{x\neq\xi_{_N}}$ and $-Gm_{_N}
 f^{_N-}_{\mu\nu}(x)_{x\neq\xi_{_N}}$, respectively,
screening these particle fluctuations. A dual approach to the nonlocal
 particle notion, based on the extended microscopic and point macroscopic
  states, will allow us to exclude point vertices from the microscopic
  Maxwell-Lorentz  and Einstein - type equations, but  to hold point
  particles in the macroscopic equations.
Our microscopic equations will manifest the local mass/charge/energy
 screening of the extended particle state by the induced field at every
  vacuum point, while our macroscopic equations will coincide with Maxwell's
   equations and reveal the origin of the scalar Ricci curvature in the
   Einstein equation.

It is well known that retarded vector (spin-1) mediators can lead only to the
 repulsion of identical sources. But the extended, nonlocal particles allow
 us to employ advanced field potentials for the gravitational attraction of
 masses-outlets in the vector field theory. In other words, we relate the
 retarded potentials $q_{_N}a^{+}_{_N\mu}$ with outgoing electromagnetic
 fields and advanced potentials
$-Gm_{_N}a^{-}_{_N\mu}$ with incoming gravitational fields in the
Maxwell-type equations. These
 elementary potentials contribute to the total gravitational, 
$B_{\mu} = - \sum_{_K}^{all} Gm_{_K}a^{-}_{_K\mu}$, and electromagnetic,
 $A_{\mu} = \sum_{_K}^{all} q_{_K}a^{+}_{_K\mu}$,
field potentials.

The unified vector structure of gravitational and electromagnetic forces,
 acting on the particle, allows to generalize the equivalence principle on
  the proper canonical four-space. The geodesic equation, $DP_{_N\mu}
  (\xi_{_N})=0$, for the charged vertex incorporates the retarded Lorentz
   and the advanced gravitational forces. The vertex charge $q_{_N}$ is
   always a source of the outgoing electromagnetic field, while the vertex
   mass $m_{_N}$ is always an outlet of the incoming gravitational field.
   The outgoing (from the point vertex) electromagnetic wave and the incoming
    (from infinity) gravitational wave are coupled vector anti-waves of each
    other. These waves never cross at finite times and they are not relevant
     to metric modulations of three-space, prohibited in the present scheme.
      The strict spatial flatness of the Universe on micro-, macro-, and
      mega-scales is the principle point of our relativistic scheme.

The global superposition of all microscopic particle states 
is considered in the last section, where we derive the macroscopic 
gravitational equations. The scalar Ricci curvature is absent in the
microscopic equations, but appears naturally
in the Einstein macroscopic equation in agreement with the Rainich-Misner
 criterion. The known criterion of double unification, the particle with its
 fields and gravity with electrodynamics, is also satisfied in the developed
  theory with nonlocal elementary objects.

\bigskip
\bigskip
\noindent {\bf 2. Flat three-space and absolute time} 

\bigskip

In order to verify the mathematical opportunity to implicate 
the flat 3D subspace $x_{_N}^i$ into the proper four-space $x^\mu_{_N}$
 with curved pseudo-Riemannian
metric under consideration of one elementary mass N, we employ the known
tetrad formalism,
 for example [5,6], which represents a proper four-interval,
  $ds_{_N}^2 \equiv g^{_N}_{\mu \nu }dx_{_N}^\mu dx_{_N}^\nu $ $\equiv$ 
$ds^2 \equiv g_{\mu \nu }dx^\mu dx^\nu $
 $\equiv \eta_{\alpha \beta  }e^\alpha_{\ \mu} e^\beta_{\ \nu} dx^\mu
  dx^\nu $ $\equiv \eta_{\alpha \beta}dx^\alpha dx^\beta $,
in the "plane" coordinates
 $dx^\alpha
\equiv e_{\ \mu} ^\alpha dx^\mu $, $dx^\beta
\equiv e_{\ \nu} ^\beta dx^\nu $,
 $\eta_{\alpha \beta }$ 
= $diag (+1,-1,-1,-1)$. One can promptly find
$e^o_{\ \mu} = \{  {\sqrt {g_{oo}}} ; - {\sqrt {g_{oo}}}g_i  \}$
and $e^a _{\ \mu} = \{0, e_{\ i}^a   \} $ from the equality
$ds^2 \equiv [{\sqrt {g_{oo}}} (dx^o-g_idx^i)]^2 - \gamma _{ij}dx^idx^j$,
$g_i \equiv - g_{oi}/g_{oo}$.
At first glance the proper spatial triad $e_{\ i}^a$ $\equiv$ $e_{_N i}^a$
 (a = 1,2,3, while $\alpha$ = 0,1,2,3) depends essentially on gravitational
  fields of external masses because this triad can be   represented via all
   components of the proper metric tensor  $ g_{oi}g_{oj}g^{-1}_{oo}
 - g_{ij}$ $\equiv$ $\gamma_{ij}\equiv \gamma^{_N}_{ij}$.
But this is not a case if there are intrinsic 
symmetries of the proper metric tensor $g^{_N}_{\mu\nu}$.

Let us consider space components $V_i$ of the metric-velocity four-vector
$V_\mu \equiv
 g_{\mu \nu } {d x}^\nu/ ds $
 by
using the  tetrad formalism, $-({\sqrt {g_{oo}  }}g_i + v_i  )(1 -
v_iv^i)^{-1/2} $
 = $V_i = e^\alpha _{\ i} V_\alpha = e^o_{\ i} V_o + e_{\ i}^a V_a $
= $-({\sqrt {g_{oo}  }}g_i + e_{\ i}^av_a  )(1 - v_av^a)^{-1/2} $.
This leads immediately to trivial relations  $v_i \equiv e_{\ i}^av_a =
 \delta _i^av_a$ between the "curved",  $v_i = \gamma_{ij}dx^j /{\sqrt
 {g_{oo}}}|dx^o - g_i dx^i|$,
and the "plane", $v_a$, three-velocities because $
e^o_{\ i} =  - {\sqrt {g_{oo}}}g_i $ and $V_\alpha = \{
(1-v_av^a)^{-1/2}; - v_a(1-v_av^a)^{-1/2} \}$.
These relations through the universal  Kronecker delta-symbols $\delta _i^a$
clearly indicate that all spatial triads
 and, consequently, the three-space metric
tensor  are  irrelevant to gravitation
 fields, {\it i.e.} $e_{_N i}^a = \delta _i^a$ and 
$\gamma^{_N}_{ij}$ = $\delta _{ij}$. Below we introduce the metric tensor
$g^{_N}_{\mu\nu}$ which satisfy these symmetries in all field potentials
  and their gauges.  

Again, Euclidean spatial geometry takes place in the 
 covariant
formalism for gravitation due to the intrinsic metric symmetries,
$g_{oi}g_{oj}g^{-1}_{oo} - g_{ij}$  $\equiv$  $ \delta_{ij}$,
of  proper four-spaces of elementary particles.
 The four-space metric tensor
in the most general case reads
$ g_{\mu \nu } \equiv
\eta_{\alpha \beta  } e^\alpha _{\ \mu} e^\beta _{\ \nu} \equiv
 \eta_{\mu \nu } +
 \eta_{\alpha \beta  } (e^\alpha _{\ \mu} e^\beta _{\ \nu} -
 \delta _\mu ^\alpha
 \delta _\nu ^\beta )$,
where  $e^o_{\ \mu} = \{  {\sqrt {g_{oo}}} ; - {\sqrt {g_{oo}}}g_i  \}$
and $e^a _{\ \mu} = \{0, \delta _i^a   \} $ $\equiv \delta_\mu^a $.
In agreement
with this consideration,  all three-intervals $dl_{_K}$ are
always associated with the universal Euclidean metrics, because
$\gamma^{_K}_{ij} \equiv g^{_K}_{oi}g^{_K}_{oj}(g^{_K}_{oo})^{-1} -
g^{_K}_{ij}
\equiv  \delta _{ij} \equiv - \eta_{ij}$ for all particles, while different 
proper four-intervals $ds_{_K}$ represent different pseudo-Riemannian metrics,
because $g^{_N}_{\mu\nu} \neq g^{_K}_{\mu\nu}$.

A proper four-interval 
of one selected mass N ($ds_{_N} \equiv ds$ and $dx_{_N}\equiv dx$, for 
short) is  given by
\begin {equation}
ds^2 = \left ( {\sqrt {g_{oo}}} { dx^o}  + {{g_{oi}dx^i }
\over  {\sqrt {g_{oo}}} }    \right )^2
 - \delta_{i j} dx^i dx^j \equiv [d {{\tau}}(s)]^2  - dl^2 
\end {equation}
 in arbitrary  external gravitational fields.
It is worth noting that (1) is a very complicated nonlinear equation with
respect to the proper metrics $ds$.
 The proper time $d\tau \equiv d\tau_{_N}(s_{_N}) \equiv {\sqrt {g_{oo}}}
 |dx^o - g_i dx^i|$  in (1) depends
on  $|ds| = {\sqrt {[d\tau (s)]^2 - dl^2}}$, and 
the nonlinear four-interval  $ds$ depends on 
 the 3D interval $dl \equiv dl_{_N} \equiv {\sqrt { \delta _{ij}dx^idx^j
  }} > 0 $ in a nontrivial way for moving bodies.

The co-ordinate time rate $dx^o_{_N}$ of the advanced
and retarded  field matter has opposite directions resulting to incoming
 and outgoing material fields. We can use the absolute time interval,
 $dt \equiv dt_{_N} \equiv {\sqrt { \gamma^{_N}_{oo}
dx_{_N}^o dx_{_N}^o  }}$  $\equiv {\sqrt { \delta _{oo} dx_{_N}^odx_{_N}^o
  }}$  $\equiv$   $   |dx^o_{_N}| > 0$, for both outgoing (retarded) and
  incoming (advanced) fields. Both the 3D space interval $dl_{_N}$
and the 1D time interval $dt_{_N}$ are based on the universal metric
symbols, $\gamma^{_N}_{ij} = \delta_{ij}$  and $\gamma^{_N}_{oo} =
\delta_{oo}$, respectively. But there is no flat, homogeneous metrics
 $ds_{_N} \equiv$   $(sign \ dx_{_N}^o){\sqrt {g^{_N}_{\mu\nu}(x_{_N})
 dx_{_N}^\mu dx_{_N}^\nu}}$
of four-space, because $g^{_N}_{\mu\nu}(x_{_N})\neq const$. The absolute
 or flat "rulers" $dl_{_K}$ and $dt_{_K}$ are relevant in practice for
 measurements and observations, and universal (Euclidean) geometry of
 3D and 1D proper subspaces ought to stand behind the world (common)
 three space
 and the world (common) time notions. Flat space and flat time, {\it i.e. }
  the flat world space+time  $\{x^i; x^o\}$, can always be applied to all
particles and fields in the Universe, while $\{x^\mu\}$ and $ds_{_N}$  is
to be specified
for every selected particle N (or for their finite ensemble).

Now we study  the metric-velocity four-vector $V_{\mu}$.  Notice that
$V_\mu = e^\alpha_{\ \mu} V_\alpha  = (e^a_{\ \mu} V_a + e^o_{\ \mu} V_o)$
= $ (e_{\ \mu}^a V_a  + \delta^o_{\mu} V_o ) $
+ $ (e_{\ \mu}^o - \delta^o_{\mu} ) V_o  $
$\equiv$ ${\cal V}_\mu + U_\mu$,  with the proper four-velocity   
$ {\cal V}_\mu  $ $\equiv$
$(e_{\ \mu}^a V_a  + \delta^o_\mu V_o) = \delta_\mu^\alpha V_\alpha$,
 because $e^a_{\ o} = 0$ and $e^a_{\ i} = \delta^a_i$.
Flat three-space geometry
is just a way  to introduce the
 four-potential
$U_\mu  \equiv (e_{\ \mu} ^o   - \delta^o_\mu  )V_o $ $= B_{\mu} + 
(q_{_N}/m_{_N})A_\mu + \partial_\mu \phi_{_N}$
 for both gravitational, $ B_{\mu}$, and electromagnetic, $(q_{_N}/
 m_{_N})A_\mu$, components.
 Let a total (canonical) particle four-momentum $P_{_N\mu }(x)$ has
  formally the inertial, gravitational and
electromagnetic contributions,
  $$P_{_N\mu }(x)\!\equiv\!m_{_N}{{g^{_N}_{\mu\nu}dx^\nu_{_N}}\over
   ds_{_N}}\!=\!\left \{\!{{m_{_N}
  }\over {\sqrt {1 - v^2}}}+{{m_{_N} ({\sqrt {g_{oo}}}-1)
  }\over {\sqrt {1-v^2}}};\!-{{m_{_N} v_i}\over {\sqrt {1 - v^2}}}-
  {{m_{_N}g_i{\sqrt{g_{oo}}}
  }\over {\sqrt {1 - v^2}}}\right \}$$
\begin {equation}
 =m_{_N}V_{_N\mu}=m_{_N} ({\cal V}_\mu
+ B_\mu  + q_{_N}m_{_N}^{-1} A_\mu + \partial_\mu \phi_{_N})
=m_{_N} ({\cal V}_\mu + B_\mu^{\neq_N}
+ S^{_N}_{\mu}),
\end {equation}
where $v_i\equiv \gamma_{ij}v^j, v^2 \equiv v_iv^i$,
$|ds|$  =  $(dx_\mu dx^\mu)^{1/2}$, $dx_{\mu} = g_{\mu\nu} dx^\nu$,
$dx^\mu \equiv dx^\mu_{_N}$,
 $v^i\equiv $  $dx^i /g_{oo}^{1/2}(dx^o-g_idx^i)$;
 $ g_i$ $=$ $-g_{oi}/g_{oo}$;  $\gamma_{ij}\equiv g_ig_jg_{oo}-g_{ij}
= \delta_{ij} = - \eta_{ij}$.

Flat three-space admits a direct analogy in structures of gravitational
 and
 electromagnetic fields (which will be proved below). Namely, the
  gravitational, $B^{}_\mu(x)\equiv -
\sum _{_K}^{all}\ Gm_{_K}
 a^{-}_{_K\mu}(x)$, and the electromagnetic, $ A_\mu (x) \!\equiv$ 
$\sum _{_K}^{all}\!q_{_K}
 a^{+}_{_K\mu}(x) $, potentials in the canonical four-momentum (2) are
  based on the elementary potentials $-Gm_{_K}a^{-}_{_K\mu}(x)$
and $q_{_K}a^{+}_{_K\mu}(x)$, respectively, with coupled forming-up
pre-potentials $a^{\pm}_{_K\mu}(x)$. The latter are a gauge family 
of  advanced (-) and retarded (+) field solutions, $a^\mp_{_K\mu}
(x)$ = ${\tilde a}^\mp_{_K\mu}(x)
+ \partial_\mu \chi^\mp_{_K}(x)$, of the Maxwell-type equations (derived
below) for the
mass $m_{_K}$ and the charge $q_{_K}$, respectively.  

Finally the canonical four-momentum $P_{_N\mu}(x)$ depends on the inertial
 momentum  $m_{_N}{\cal V}_{_N\mu}(x)$, the "internal" self-momentum
 $m_{_N}S^{_N}_{\mu}(x)$ $\equiv$ $  q_{_N}
A_{\mu}(x)$ $-$ $Gm^2_{_N}a^{-}_{_N\mu}(x)$ + 
$m_{_N}\partial_\mu \phi_{_N} (x)$, and the "external"  field
momentum $m_{_N}B^{\neq_N}_\mu(x)$ $\equiv$  $- 
 m_{_N}\sum_{_K}^{_K\neq _N}\ Gm_{_K} a^{-}_{_K\mu}(x)$ in the most general
  case.

 The "internal" self-potential $S^{_N}_{\mu}(x)$
depends on the proper gauge, $\partial_\mu \phi_{_N}$, and the particle
parameters, $m_{_N}$ and $q_{_N}$, while the "external", gravitational
potential
is independent from the proper particle parameters. A physical gauge can
 assist to avoid both a self-action in the proper four-interval $ds_{_N}$
  and an infinite energy of the self-field N at the vertex point $\xi_{_N}$.

At first we study the relations (2) for an electrically uncharged particle,
when $q_{_N}=0$ and $S^{_N}_{\mu} (x) = \partial_\mu \phi_{_N} - Gm_{_N}
a^{-}_{_N\mu}(x)$.
Notice  from the relations (2) that the passive gravitational mass in
$m_{_N}B^{\neq_N}_\mu$ is
always equal to the inertial (rest) mass in $m_{_N}{\cal V}_\mu$. 
Contrary to General Relativity, the electrodynamics-like gravitation
(based on the linear superposition of elementary fields) 
  may propose from (2) a detail structure of the proper metric
tensor $g_{\mu\nu}^{_N}(x)$. The proper tetrad  takes  the following
components
$e_{\ \mu} ^a = \{0,   \delta_i^a \}$
$= \delta^a_\mu$ and $e^o_{\ \mu} =
\{1 +   {\sqrt {1-v^2} } (B_o + \partial_o \phi_{_N});
 {\sqrt {1-v^2} }(B_i + \partial_o \phi_{_N}) \} $
 $= \delta^o_\mu +  {\sqrt {1-v^2} }(B_\mu + \partial_\mu \phi_{_N})   $.
  Than the proper  metric tensor $g^{_N}_{\mu \nu } \equiv g^{_N}_{\mu\nu}
  (x_{_N})$
for the selected mass $m_{_N}$  is given  by
\begin{equation}
\cases  { g^{_N}_{oo} = [1 +   {\sqrt {1-v^2} }(B^{\neq _N}_o + S^{_N}_o)
 ]^2   \cr
       g^{_N}_{oi} = [1 +   {\sqrt {1-v^2} }(B^{\neq _N}_o + S^{_N}_o)]
 {\sqrt {1-v^2} }(B^{\neq _N}_i + S^{_N}_i)    \cr
 g^{_N}_{ij} = ({{1-v^2} })(B^{\neq _N}_i + S^{_N}_i)(B^{\neq _N}_j +
  S^{_N}_j)   - \delta_{ij}   \cr }.
\end{equation}

Notice that the proper gravitational potential, ${\cal B}^{_N}_\mu (x)
\equiv - Gm_{_N}a^{-}_{_N\mu}(x) \neq 0$ in (3), does not vanish in the
absence of external gravitational fields, when $B^{\neq _N}_\mu = 0$. This
 proper potential should take the Newtonian form in the particle rest frame,
  with $B^{_N}_o(x)=
|{\bf x}-{\mbox {\boldmath $\xi $} }_{_N}|^{-1} \rightarrow
\infty$ when $|{\bf x}-{\mbox {\boldmath $\xi $} }_{_N}| \rightarrow 0$.  An
infinite value of such proper potential, ${\cal A}^{_N}_\mu (x) \equiv
q_{_N}a^{+}_{_N\mu}(x)$, of the point electric charge
$q_{_N}$ at its path points $\xi_{_N}$ is conventionally omitted in
 Classical
Electrodynamics.
The similar divergence problem may be resolved more appropriately in
(2)-(3) due to a physical choice of the gauge term $\partial_\mu \phi_{_N}$,
 when $S^{_N}_{\mu}(x)P_{_N}^\mu(x) $ = $[\partial_\mu \phi_{_N} -
 Gm_{_N} a^{-}_{_N\mu}(x, \xi_{_N})] P_{_N}^\mu (x)$ = 0. In other words,
the internal self-momentum behaves like an internal angular momentum
or a particle spin with
$S^{_N}_{\mu} (x)= \{0; {\bf S}(x) \}$ in the rest frame of references.

As it was expected, all components of the three-space metric
 tensor  are independent from gravitational fields. Now the equality
  $\gamma_{ij} \equiv g_{oi}g_{oj}g_{oo}^{-1} - g_{ij} = \delta _{ij}$
may be verified  directly from (3) despite every component of the  
metric tensor $g_{\mu\nu}$ depends on the field potentials. 
One can represent (3) in a more compact way, 
$g_{\mu\nu} = \eta_{\alpha\beta}e^\alpha_{\ \mu} e^\beta_{\ \nu}$
and    $ e_{\ \mu}^\alpha = \delta^\alpha_\mu +
 \delta^{\alpha o} {\sqrt {1-v^2} }B'_\mu    $.
Notice, that 
$P_{_N}^{\mu} = g^{\mu\nu}P_{_N\nu}$ = $m_{_N}[\eta^{\mu\nu}{\cal V}_\nu
 + (g^{\mu\nu} -
\eta^{\mu\nu}){\cal V}_\nu + g^{\mu\nu}B'_\nu  ]$
= $m_{_N}\eta^{\mu\nu}{\cal V}_\nu$ - $m_{_N} \{ (1 + {\sqrt {1-v^2}}B'_o
  )^{-1}
(B'_o + B'_i v^i); 0 \}$ = $m_{_N}(1-v^2)^{-1/2} \{1 + 
(g_{oo}^{-1/2} - 1 + g_i v^i  ); v^i \}  $ and $P_{_N\mu}P_{_N}^{\mu}$ = 
$m^2_{_N} ({\cal V}_\mu {\cal V}^\mu  + 0)$ = $m^2_{_N}$ for an arbitrary 
gravitational four-potential $B'_{_N} \equiv $  $B_\mu^{\neq_N}+
S^{_N}_\mu$ $ = B_\mu + \partial_\mu \phi_{_N} $, when $q_{_N}$ = 0.

Substituting the metric tensor (3) into  (1), we
can obtain a general equation for the proper four-interval,   $ds = ds_{_N}$,
 of one selected mass $m_{_N}$,
\begin {equation}
 ds^2 + dl^2 =  (d{x}^o + {\sqrt{1-v^2}}
B'_\mu V^\mu ds)^2,
\end {equation}
with $ dl^2  \equiv \delta _{ij}dx^idx^j  $.
From here the proper particle time $d\tau$ and the particle four-interval
$ds$ are not self-affected by the particle mass and charge only under the
 physical gauge $S^{_N}_{\mu}P_{_N}^\mu$
 $\equiv $  $m_{_N}S^{_N}_{\mu}V_{_N}^\mu$ = 0,
when $B'_\mu{V}^\mu = (B_{\mu}^{\neq_N} + S^{_N}_{\mu})V^\mu  $
= $B_{\mu}^{\neq_N} V^\mu$.

Each elementary gravitational potential $-Gm_{_K}a^{-}_{_K\mu }(x)$ 
in $B^{\neq_N}_\mu(x)$ is related to
its distant point vertex $\xi_{_K}$ in agreement with  the  Maxwell-type
equation (derived below). This potential in the two-body problem has only
 Newton's component,
$B^{\neq_N}_\mu = $ $-Gm_{_K}a^{-}_{_K\mu }(x)$ =  $\{-Gm_{_K}  |{\bf x}-
{\mbox {\boldmath $\xi $} }_{_K}|^{-1} ; 0 \}$, for a point mass $m_{_N}$
 at $x=\xi_{_N}$ in the mass $m_{_K}$ frame of references. Below we
use this potential to study gravitational phenomena in the Sun's  field.

\bigskip
 \bigskip
\noindent {\bf 3. Relativity tests in the Sun's field} 

\bigskip

\noindent {\it 3.1. The planetary perihelion precession in the flat world
 space}
 
\bigskip

Now we derive the
planetary perihelion precession in order to test in practice the
four-interval equations (4)  with the novel structure
of the metric tensor (3), which is consistent with flat three-space,
 $\gamma _{ij} = \delta _{ij}$.
A distant system of elementary gravitational centres K may be considered
 as one united
center with the joint mass 
(the Sun, for example, with the Sun's mass $M_{_S}$),
 when all $|{\bf x}-{\mbox {\boldmath $\xi $} }_{_K}| \approx r \equiv
  u^{-1}  $. Nonlinear relativistic relations for the function $\alpha_{_N}
  (ds^2,dl^2) $=$- {\sqrt{1-v_{_N}^2}}
B^{\neq_N}_\mu V_{_N}^\mu $ will be derived in the next section. 
One may use in (4) the Newton potential, 
$(-B^{\neq_N}_o) =  GM_{_S}r^{-1} \equiv \mu u \ll 1$,  for the
non-relativistic motion when a  considered mass N  (a planet
with  mass $m_{_N} \ll M_{_S}$ and  $v_{_N}^2 \equiv dl^2/d\tau^2 \ll 1$)
 moves in the Sun's
rest frame, where $B^{\neq_N}_{i} = 0$.
The equation (4) for the proper four-interval $ds$ reads 
\begin{eqnarray}
 ds^2(t,l,\tau) + dl^2  \equiv d\tau^2(t,l,s) =
dt^2 \left(1 -\mu u{\sqrt {1 - {{dl^2}\over d\tau^2} }}   \right)^2
\nonumber \\
\approx dt^2 \left ( (1-\mu u)^2 + \mu u (1-\mu u){{dl^2}\over d\tau^2}
 \right )
\approx (1-2\mu u)dt^2 +  \mu u dl^2, 
\end{eqnarray}
where we used   $dt \equiv |dx^o|$,  $\mu u \ll 1$,  $dl^2   \ll    d\tau^2$,
 and $dt^2-d\tau^2 \ll dt^2$.

The mass dependent coefficient $1 - \mu u$ in (5) at
the flat three-interval, $dl^2 = dr^2 + r^2d\varphi^2 $
= $u^{-4}du^2 + u^{-2}d\varphi^2$, does not mean
departure from Euclidean  spatial geometry
in gravitational fields. This coefficient is related to a direct involvement
 of the space replacement $dl$ into the
proper time $d\tau (dt,dl,ds)$. Our proper time in (5),
$d\tau \approx {{(1 - 2\mu u + \mu u v^2 )^{1/2}} }dt$,  coincides with
the General Relativity proper time, ${{(1-2\mu u)^{1/2}} }dt$,
in the limit of weak fields and small velocities. 

The Killing vectors and integrals of motion,
$(1 - \mu u)^{2} {d t / ds} = E = const   $ and
 $r^2{d \varphi / ds} = L =const$ (with $\vartheta = \pi /2 = const$),
  are well known
under  the four-interval  with stationary
coefficients, for example [7]. By taking into account
these conservation laws in  (5), one obtains
an equation for a rosette motion of planets
under the non-relativistic  restrictions  on their velocities,
\begin{equation}
{(1-2\mu u) L^{-2}} + {(1-3\mu u)({u'^2 + u^2})} = {{E^2}L^{-2}},
\end{equation}
where $u'\equiv du/d\varphi$ and $\mu u \ll 1$. Now one may
differentiate (6) with respect to
the polar angle
$\varphi$,
\begin{equation}
u'' + u - {\mu\over L^2 } = {9\over 2} \mu  u^2 +
3\mu u'' u + {3\over 2}\mu  u'^2,
\end{equation}
by keeping only the oldest nonlinear terms. This equation may be
 solved in two steps, when a linear
 solution, $u_o = \mu L^{-2}(1 + \epsilon cos \varphi)  $,
 is substituted  into the nonlinear terms
 on the right hand side of (7).

The most important
correction (which  is summed over century
rotations of the planets) is related to the "resonance"
(proportional to $ \epsilon cos \varphi$) nonlinear terms.
 We therefore ignore
in (7) all corrections, apart from $u^2 \sim 2\mu ^2L^{-4}
\epsilon cos \varphi$
and $u'' u \sim - \mu ^2L^{-4}\epsilon cos \varphi$. Then the  approximate
equation for the rosette motion, $u'' + u - \mu L^{-2} \approx
6\mu ^3 L^{-4} \epsilon cos \varphi  $, leads   to the
well known perihelion precession, $\Delta \varphi = 6\pi \mu ^2L^{-2}
\equiv 6\pi \mu / a (1-\epsilon ^2) $, which also may be derived through
Schwarzschild's metric formalism with curved three-space, for example
[5-10]. 

It is important to emphasize for a verification of our space concept
 that the
 observed result  for the planet 
 perihelion precession  $\Delta \varphi$ in the Sun's field
was derived  from the GR nonlinear four-interval (5) under  flat
three-space, rather than
from the linear four-interval under curved three-space.

\bigskip
	
\noindent {\it 3.2. The radar echo delay in flat three-space}

\bigskip
The gravitational redshift of light frequency $\omega$ is still considered
 in some textbooks as a confirmation of the accepted opinion that gravity
 couples to the energy content
of any matter, including the photon's energy $E_\gamma$ or the 
"relativistic mass" $m_\gamma = E_\gamma/c^2$.
We agree completely with the statement $E=mc^2$ for rest-mass particles,
but disagree with its inverse reading, $m = E/c^2$, for all kinds of
matter, in particular for photons with $m = 0$ and $E\neq 0$.   

 In 1907 Einstein 
introduced the principle of equivalence for a uniformly accelerated body
 and concluded that its potential energy is associated with the "heavy"
 (passive) gravitational mass [11]. This Einstein's conclusion was formally
 generalized in a way that any energy,
including light, has a gravitational mass.  
Proponents of this generalization assume that photon's energy-"relativistic
 mass" is attracted directly by the Earth in agreement with
the measured redshift  $\Delta \omega / \omega =  \Delta E_\gamma / E_\gamma
 = \Delta (-m_\gamma GM_{_E}R_{_E}^{-1}) / m_\gamma c^2=-\Delta GM_{_E}/
  R_{_E} c^2 $ due to these formal relations for light in the static
   gravitational field.
But the formal introduction of the "relativistic mass" for matter with
 a zero scalar mass invariant leads to the underestimated light deflection,
  $\varphi = - 2GM_{_S}/R_{_S}c^2 \equiv - 2\mu/c^2R_{_S} \equiv - r_g/R_{_S} $,
   under the Newtonian "fall" of photons in the Sun's gravitational field [12].
    In 1917, when the Schwarzchild's
solution [2] for space curvature had been accepted by General Relativity,
 Einstein predicted the non-Newtonian light deflection $\varphi = - 2r_g/R_{_S}
  $, based once again on a direct attraction of the photon's
energy $E_\gamma$ by the Sun. All 
 experimental tests  supported later the corrected
Einstein's result for the gravitational light deflection, that up till now
holds non-Euclidean three-space in  contemporary developments of General
Relativity.

But do photon's gravitational phenomena undoubtedly confirm that space is
 really curved and gravity couples  to the photon's energy?
The principle of equivalence was proposed only for a rest-mass body  with
 a proper
reference system [11], while the photon has no inertial or  rest mass
at all. How
may a passive gravitational mass be attributed to 
a particle without an inertial mass?

Now we revise the conventional theory of the
radar echo delay  and the gravitational 
light deflection. Our goal is to show that
 the known measurements [9,13,14] of light phenomena in the Solar system 
may be interpreted as confirmations of the
world space flatness. Let us consider a static ($g_i = 0$, for simplicity)
 gravitation field, where the physical slowness of photons, $n^{-1} \equiv
  v/c $,
can be derived directly from the covariant Maxwell equations [5], $n^{-1} 
= {\sqrt  {{\tilde \epsilon} {\tilde \mu}} } =  {\sqrt {g_{oo}} }$.
The measured, physical  velocity $v = dl/d\tau$,
as well as the measured, physical  frequency $\omega = \omega_o dt/d\tau$,
 is
to be specified  with respect to the  physical time  $d\tau = {\sqrt {g_{oo}}
 }dt$.
 Originally  Einstein associated 
the light's redshift with the different clock rates in the Sun's
gravitational potential [11], and this true nature of the redshift
 is irrelevant to the energy content or to the "apparent weight" [13] of the
  massless photon. In other words, the light redshift is simply the
  gravitational blue shift of clocks which measure the photon frequency.

As compared to the physical velocity of light, its co-ordinate velocity
 $v = dl/dt$ is double shifted by the  static gravitational potential
${\sqrt {g_{oo}} }$,  
\begin {equation}
{\dot l} \equiv {{dl}\over dt} \equiv {{dl}\over d\tau}{{d\tau}\over dt}
= c n^{-1} {\sqrt {g_{oo}} }  = c{{g_{oo}}} = c \left(1 
 - {{r_g}\over 2r}\right)^2, 
\end {equation}
 in the Sun's gravitational field, where $r_g \equiv 2GM_{_S}/c^{2} =
 2,95$ km,  $r_{g}/r \ll 1 $.   Notice that the local physical slowness
  $n^{-1} = {\sqrt {g_{oo}}} $, and the local time dilation $ d\tau / dt
   \equiv {\sqrt {g_{oo}}}
|1 - g_i{\dot x}^i| \approx {\sqrt {g_{oo}}}$ are responsible together for
 the double co-ordinate slowness in the relation (8).

A world time delay of Mercury's radar echo reads through the co-ordinate
relation (8) as
\begin {equation}
\Delta t = {2} \int_{l_{_E}}^{l_{_M}} dl \left ( 
{{1}\over {\dot l}} - {1\over c} \right ) \approx 
{2\over c} \int_{x_{_E}}^{x_{_M}}  
  {{r_g dx } \over {\sqrt {x^2 + R_{_S}^2} }  }   
\approx {2r_g\over c } ln {{ 4r_{_M{_S}} r_{_E{_S}} }\over R^2_{_S} } =
220\mu s,
\end {equation}
where $y \approx R_{_S} = 0.7\times 10^6$ km is the radius of the Sun,
while
$r_{_E{_S}} = 149.5\times 10^6$km and $r_{_M{_S}} = 57.9\times 10^6$km
are the Earth-Sun and Mercury-Sun distances, respectively. It is essential
that we use Euclidean metrics for any finite distance, $r = (x^2 +
y^2)^{1/2}$, between the Sun's center (0,0)
and any considered point (x, y) on the photonic axis x.       
One can measure in the Earth's laboratory only the physical time delay
 $\Delta \tau_{_E}$, which practically coincides with the absolute world
  time delay $\Delta t$ in the
Earth's  weak field, {\it i.e.}  $\Delta \tau_{_E} \approx \Delta
t = 220 \mu s$. The known experimental results [14] correspond to the radar
echo delay (9), based on the concept of flat world space.

\bigskip

\noindent {\it 3.3. The gravitational light bending in flat three-space}

\bigskip
A co-ordinate angular deflection $\varphi$ of a light front in
the Sun's gravitational field may be derived geometrically by using the
 co-ordinate velocity (8),

\begin {eqnarray}
\varphi_\infty  = - 2\int_{0}^{\infty} dl {\partial\over
\partial y }
\left ( {{\dot l}\over c} \right ) \approx - 2\int_{0}^{\infty} dx
{\partial\over
\partial y } \left ( {r_g\over
{\sqrt {x^2 + y^2}}}\right )
\nonumber \\
 \approx -2r_g\int_0^\infty dx {{R_{_S}}\over (x^2 + R_{_S}^2)^{3/2}} = -
 {{2r_g}\over R_{_S}} = - 1.75".
\end {eqnarray}

A physical way to derive the ray  deflection (10) is to apply Fermat's
principle to light in gravitational fields.  We 
relate the covariant wave vector $K_{o}$ in the scalar equation 
$K_\mu K^\mu =  0$ to the measured, physical energy-frequency  of the photon
 ($K_o = E = \hbar\omega = \hbar\omega_o / {\sqrt {g_{oo}}}, \ \hbar\omega_o
  = const > 0$; notice that $P_o$  is also the measured particle's energy
  in $P_\mu P^\mu = m^2$) in agreement with the original  Einstein's paper
   [11].

The scalar wave equation $K_\mu K^\mu = g_{_N}^{\mu\nu}K_\mu K_\nu =  0$
has
two solutions for paired anti-waves. One solution for the electromagnetic
 wave (photon), with
 \begin {equation}
{\cases { K_o \equiv \hbar \omega_o dt/ c d\tau  =
 {{g_{oo}}} (K^o - g_i K^i)  \cr
\gamma_{ij}K^iK^j=g_{oo}(K^o - g_i K^i)^2=K^2_o/g_{oo}=\hbar^2\omega^2_o
dt^2 /c^2g_{oo} d\tau^2 \cr
K^i = \hbar \omega_o dt dx^i / c {\sqrt {g_{oo}}} d\tau dl,
 K_i =
 - (\hbar\omega_o dt/ c {\sqrt {g_{oo}}} d\tau) [(\gamma_{ij}dx^j / dl) +
  {\sqrt {g_{oo}}}g_i ]\cr }},
\end {equation} 
or $K_\mu = \{+E, \mp E[(\delta_{ij}dx^i/{\sqrt {g_{oo}} dl} ) + g_i] \}$,
and another solution for the gravitational wave (graviton) $G_\mu \equiv
 - K_{_\mu} = \{-E, \pm E[(\delta_{ij}dx^i/{\sqrt {g_{oo}} dl} ) + g_i] \}$,
 with $G_\mu G^\nu = 0$.
The graviton, the anti-photon in our consideration, is associated formally
with the negative energy $G_o = -E < 0$
(like the negative-energy right-handed neutrino in the Weyl equation)
or with the backward (advanced) wave motion under the positive world time
direction, $dt = |dx^o| > 0$.  By employing one absolute time rate $dt > 0$
 for measurements, one should relate gravity with the advanced, incoming
 material fields moving from infinity to their point outlets $\xi^i_{_K}$
  in flat world space $x^i$.

The Fermat variations with respect to $\delta \varphi$ and $\delta u$ 
($ r \equiv u^{-1} $, $\varphi$, and  $\vartheta = \pi/2$ 
are the spherical coordinates) for photons in a static  gravitational field,
\begin {equation}
\delta\!\int K_i dx^i = -\delta\!\int\! {{\hbar \omega_o\gamma_{ij}dx^j}
\over
c g_{oo}dl}dx^i =   -{\hbar\omega_o\over c}\delta\!\int  {  { \sqrt {
  du^2 + u^2d\varphi^2   }   }\over u^2 (1- 2^{-1}r_gu)^2  } = 0,
\end {equation}
where $\gamma_{ij} = \delta_{ij}$ and  $dl = {\sqrt { \delta_{ij} x^ix^j}}$
  = $
{\sqrt {       dr^2 + r^2d\varphi^2}}$ is the Euclidean three-interval,
leads to a couple of light ray equations,
\begin {equation}
{\cases {
(1- 2^{-1}r_g u)^4 \left [ \left(  {u'_\varphi} \right )^2 +
u^2    \right] = U_o^2 = const \cr
 u''_{\varphi\varphi} + u =  U_o^2 r_g (1 - 2^{-1}r_g u)^{-5} \approx
 U_o^2 r_g  \cr}}.
\end {equation}

Solutions of (13),
 $u \equiv r^{-1}$ $ = r^{-1}_o sin \varphi $ + $r_g r^{-2}_o 
(1 + cos \varphi)$ and $r_g/r_o \approx r_gU_o \approx r_g/R_{S} \ll 1 $,
 may be used under the Sun's weak field. The propagation of  light from
 $r(-\infty) = \infty, \varphi (-\infty)= \pi$
to  $r (+\infty)  \rightarrow  \infty , \varphi (+\infty)
 \rightarrow \varphi_{\infty} $ corresponds to the angular deflection 
 $ \varphi_{\infty} = arsin [-2r_g R_{_S}^{-1}(1 + cos\varphi_\infty)]$
  $\approx -2r_g/R_{_S}$ from the initial light's direction.
This deflection coincides with (10)
and is in agreement with the known measurements $-1,66''  \pm 0.18'' $,
 for example [9].

The massless photon is not
attracted directly by the Sun, but exhibits the physical velocity slowness
 $v/c \equiv
dl/c d\tau = {\sqrt
{g_{oo}}} $, with $g_{oo} < 1$
and $\gamma_{ij} = \delta_{ij}$, under locally dilated time rate. Both the
 photon and the graviton
have neither inertial nor gravitational masses. Massless gravitons are not
 coupled to each other, and  gravitational interactions, like electrodynamic
  ones, are linear in the present theory of the flat world space with the
  absolute time rate.
The mass four-current is the origin of the vector gravitational field
 according to the Maxwell-type equation (derived below). This nature of
  gravity is in agreement with the zero measurements of the Nordtvedt
  effect [15,16] and satisfies the Einstein principle of equivalence. Both
  the vector gravitational wave (or graviton) and the vector electromagnetic
   wave
 (or photon) have zero mass and charge four-currents, respectively. These
  waves can not generate themselves gravitational and electromagnetic
  fields, and our gravitons, like photons, do not couple to each other.

The numerical results (9) and (10) are well known and were proposed by
 many authors, but we have derived these results for light in the flat
  world space.
Accepting the verified agreement of 
 $\Delta \varphi$ from (7),  $\Delta t$ from (9),
and   $ \varphi_\infty $ from (10) with the relevant measurements, one
 may conclude that the Mercury perihelion precession, the radar echo delay,
 and the gravitational light deflection by the Sun reinforce the credibility
  of the flat Universe.

\bigskip \bigskip

\noindent {\bf 4. Proper four-space for the charged mass}

\bigskip

The observable evolution of matter is three-dimensional in spite
 of the fact that the proper four-space or other high dimensional manifolds
 can be employed for
a self-consistent description of any selected mass. 
Geometry 
of proper high dimensional manifolds may differ from Euclidean
 geometry of their 3D intersection called the world space.
 This provides an opportunity
 to  implicate  electric charges into the metric tensor of 
the formally introduced pseudo-Riemannian four-space $x = x_{_N}$.

In order to describe one selected mass $m_{_N}$ with the electric charge
$q_{_N}$ we employed  the symmetrical involvement of masses and charges
into the   proper particle
four-momentum  (2) or the total four-velocity $V_{_N\mu} \equiv
m^{-1}_{_N}P_{_N\mu}$,
\begin {eqnarray}  V_{_N\mu}(x) = 
  {\cal V}_\mu  + \sum_{_K}^{all}
[-Gm_{_K}a^{-}_{_K\mu }(x) + m_{_N}^{-1}q_{_N}q_{_K}a^{+}_{_K\mu } 
 (x)] + \partial_\mu \phi_{_N}(x)
\nonumber \\
 \equiv  {\cal V}_\mu 
+ U_\mu(x)
= \left \{  {1\over {\sqrt {1 - v^2}}  }; 
 - {{ \delta_{ij}v^j} \over {\sqrt {1 - v^2}}  } \right \}    +
     \left \{ {{ ({\sqrt {g_{oo}}}-1)
  }\over {\sqrt {1-v^2}}};
-{{g_i{\sqrt{g_{oo}}}
  }\over {\sqrt {1 - v^2}}}\right \}, 
\end {eqnarray}
 where $U_\mu(x)
 $  $ \equiv   \sum_{_K}^{all}
[ -G m_{_K}a^{-}_{_K\mu }
 (x) + m^{-1}_{_N}q_{_N}q_{_K}a^{+}_{_K\mu }
 (x)] + \partial_\mu \phi_{_N}$ 
 $\equiv$ $B^{\neq_N}_\mu (x)$ $ + $ $S^{_N}_{\mu} (x)$, 
$S^{_N}_{\mu} (x)$ $\equiv $ $  -
G m_{_N}a^{-}_{_N\mu} (x)  $   $ + \partial_\mu \phi_{_N}(x)$
 $+  m_{_N}^{-1}q_{_N}\sum_{_K}^{all} q_{_K}a^{+}_{_K\mu }
 (x)$,
$ {\cal V}_\mu (x) \equiv \delta ^\alpha _\mu V_\alpha $ = $ \{
\beta^{-1}, - \beta^{-1}\delta_{ij}v^j \}$,
and $\beta$=$|ds|/d\tau$=$|ds|/(ds^2 + dl^2)^{-1/2}$=${\sqrt
{1-\delta_{ij}v^iv^j}}$.

By employing the coupled pre-potentials $a^{\pm}_{_K\mu }$ for the
 electromagnetic, $q_{_K} a^{+}_{_K\mu }$, and gravitational,
$-Gm_{_K}a^{-}_{_K\mu }$, field potentials,
 one can introduce for every charged mass the proper four-dimensional
  space  ${x}_{_N}^\mu$ with
 affine connections generated by 
 both external masses and electric charges.
 The total (canonical) four-momentum of the charged mass  in its proper
  pseudo-Riemannian four-space takes the "old", mechanical view,
  $P_{_N}^\mu =
m_{_N}d{x}_{_N}^\mu/ds_{_N}$, $ P_{_N\nu} = m_{_N}{g}^{_N}_{\mu\nu}
d{x}_{_N}^\mu/ds_{_N} $ = $
m_{_N} (\delta^\alpha_\nu V_\alpha + B_{\nu})
 + q_{_N}A_{\nu}$ + $ m_{_N}\partial_\mu \phi_{_N}$, 
  $ds_{_N}^2$  = ${ g}^{_N}_{\mu\nu}d{ x}_{_N}^\mu 
d{x}_{_N}^\nu $, $P_{_N\mu}P^{\mu}_{_N}$ 
  $ = m_{_N}^2$.
In principle, all charges  $q_{_K}$ and masses $m_{_K}$ 
 may contribute into the proper (canonical) metric tensor for the
  mass $m_{_N}$,
\begin{equation}
\cases  { g^{_N}_{oo} = (1 +   \beta_{_N}U^{}_o )^2   \cr
       g^{_N}_{oi} = (1 +   \beta_{_N}U^{}_o )
 \beta_{_N}U^{}_i    \cr
 g^{_N}_{ij} = \beta_{_N}^2 U^{}_i U^{}_j   - \delta_{ij}   \cr
g^{_N}_{oi}g^{_N}_{oj}(g^{_N}_{oo})^{-1} - g^{_N}_{ij} = \delta_{ij}
},
\end{equation}
into the proper (canonical) tetrad ($  {g}^{_N}_{\mu\nu} \equiv{g}_{\mu\nu}= 
\eta_{\alpha\beta}e^\alpha_{_N\mu} e^\beta_{_N\nu}$),
\begin {equation}  
e^\alpha_{_N\mu}(x)  = \delta^\alpha_{\mu}
 + \delta^{\alpha o} \beta_{_N} U^{}_\mu (x)\equiv 
\delta^\alpha_{\mu}
 + \delta^{\alpha o} \beta_{_N} (B^{\neq_N}_\mu + S_{_N\mu} )
\equiv \delta^\alpha_{\mu}
 + \delta^{\alpha o} \beta_{_N} (V_\mu - {\cal V}_\mu),
\end {equation}
and into the affine connection in the proper (canonical)
 four-space $x_{_N}$,
\begin {equation}
\Gamma_{\mu\nu}^\lambda = {1\over 2} g_{_N}^{\lambda\rho}
(\partial_\mu g^{_N}_{\nu\rho} +  \partial_\nu g^{_N}_{\mu\rho} -
\partial_\rho g^{_N}_{\mu\nu}).
\end {equation}
The point is that all potentials in (15)-(17) are not observable
notions and the gauge field ought to be selected properly in order to avoid
 nonphysical results.
One can verify from (15) that the three-space metric tensor $\gamma_{ij}
\equiv {g}_{oi}{ g}_{oj}
{g}^{-1}_{oo}  - { g}_{ij}$  = $\delta _{ij}$ corresponds to  Euclidean
3D sub-space  under any gauge
transformations. These six bounds admit only four independent functions,
$\beta_{_N} U_\mu $, for the ten different metric components in (15). And
 these four functions have three transitional degrees of freedom due to
 the particle velocity $v_{_N}^i$ in $\beta_{_N}$, four gravitational
 degrees of freedom due to the external field potential $B^{\neq_N}_\mu$,
  and three rotational degrees of freedom due to the internal self-potential
   $S_\mu^{_N}$, with $S_\mu^{_N}P_{_N}^\mu = 0$. We prove below that the
   proper particle time $d\tau_{_N}$ is not related to new degrees of
    freedom, {\it i.e.}  there are only ten degrees of freedom
in the pseudo-Riemannian metric tensor (15) for the most general case.  

The physical three-velocity of the mass N in the flat world space,
 $d{x}_{_N}^i/d\tau_{_N}$ $ \equiv v^i_{_N} \equiv v^i = \delta ^{ij} v_j $,
   is determined by the  proper time $d\tau_{_N}$.
The proper time of one selected charged particle N, 
$d\tau_{_N}\equiv \beta^{-1} |ds|$ = $|{\sqrt {g_{oo}}}(dx^o-g_idx^i)|$
 = $|(1 + \beta U_o)d{x}^o$
+ $\beta U_id{x}^i|$ = $|d{x}^o + \beta U_\mu d{x}^\mu|$ = 
$|d{x}^o + \beta^2 U_\mu
V^\mu  d\tau sign \ ds|$ and $sign \ ds = sign \ dx^o$, depends on the
 external
 gravitational potential $B_\mu^{\neq_N}$,
\begin {eqnarray}
 {{{d \tau_{_N}}}\over |dx^o| } 
\equiv {{1}\over 1 - \beta_{_N}^2(\tau_{_N}) U_\mu V^\mu_{_N}(\tau_{_N})  }
 \equiv  {  {  1 + \beta_{_N}(\tau_{_N}) U_o  }\over
 {1 - \beta_{_N} (\tau_{_N})U_i v_{_N}^i(\tau_{_N}) }}
\nonumber \\
={{1}\over 1 - \beta_{_N}^2(\tau_{_N}) B^{\neq_N}_\mu V^\mu_{_N}(\tau_{_N})
 }\equiv {  {  1 + \beta_{_N}(\tau_{_N}) B^{\neq_N}_o  }\over
 {1 - \beta_{_N} (\tau_{_N})B^{\neq_N}_i v_{_N}^i(\tau_{_N}) }}.
 \end {eqnarray}  

We used in the equation (18) the physical, spin-type gauge  with
$S^{_N}_{\mu}V_{_N}^\mu = 0$, when  $S_o(1 - \beta U_i v^i) + S_i v^i
 (1+\beta U_o)$ = $S_o(1 - \beta B^{\neq_N}_i v^i) + S_i v^i (1+\beta
  B^{\neq_N}_o)= 0$. Then the self-potential $S_{\mu}\equiv S_\mu^{_N}$
   or the particle mass $m_{_N}$ and charge $q_{_N}$
do not contribute to the proper particle time $d \tau_{_N}$ in the physical
 velocity $v_{_N}^i = dx_{_N}^i/d\tau_{_N}$.
The proper four-interval $ds_{_N}^2 = d\tau_{_N}^2 - \delta_{ij}dx^idx^j$ 
of the electrically charged particle N depends only on the
external gravitational potential $B_{\mu}^{\neq_N}(x)$ and takes
pseudo-Euclidean form in the absence of external masses.

By taking into account that  $\beta_{_N} \equiv \beta = |ds|/ (ds^2 +
 dl^2)^{1/2}$, one can read (18) as follows,
\begin {equation}
 {{{d \tau_{_N}}}\over dt } 
\equiv {1\over 2} + {\sqrt {{1\over 4} +  U_\mu V^\mu_{_N}(\tau_{_N})
 {{ds_{_N}^2}\over dt^2}    }}
= {1\over 2} + {\sqrt {{1\over 4} +  B^{\neq_N}_\mu V^\mu_{_N}(\tau_{_N})
 {{ds_{_N}^2}\over dt^2} }}.
\end {equation}

 The basic interval relation $ds^2_{_N} + dl^2_{_N} \equiv 
(dx_{_N}^o + \beta U_\mu dx^\mu )^2 \equiv
(dx_{_N}^o 
- \alpha_{_N} ds_{_N} )^2$, with $\alpha_{_N}  \equiv - \beta U_{\mu}
V^\mu$ = $- \beta B^{\neq_N}_{\mu}V^\mu$
$\equiv - (B^{\neq_N}_o + B^{\neq_N}_i v^i )/(1 + \beta B^{\neq_N}_o)$,
$ds^2_{_N}\equiv {{
g^{_N}_{\mu\nu}dx_{_N}^\mu dx_{_N}^\nu}}$, $dl^2_{_N}  \equiv {{\delta_{ij}
dx_{_N}^idx_{_N}^j}}$, and $dt^2_{_N} \equiv {{\delta_{oo} dx_{_N}^o
dx_{_N}^o}}$, may be also represented in the following form,
\begin {equation}
ds^2  \equiv
  \left ( {{{\sqrt {(dx^o)^2 - dl^2(1-\alpha_{_N}^2)} }  \mp \alpha_{_N} 
dx^o }
\over (1 - \alpha_{_N}^2)}   \right )^2
 \approx {{dt^2}\over (1+\alpha_{_N})^2}
- {{dl^2}\over (1+\alpha_{_N})},
\end {equation}  
when $(1-\alpha_{_N}^2)dl^2/dt^2 \ll 1 $. By selecting the only one sign
 in (20), we used
from (18) that $ds^2 = dt^2(1+\alpha_{_N})^{-2}$ when $dl^2 = 0$. 
Notice that $U_{\mu}V^\mu$
 $= $ $B^{\neq_N}_{\mu}V^\mu < 0$ and $\alpha_{_N} > 0$ for strong
 gravitational fields,
{\it i.e.}  
there is no Schwarzschild-type peculiarity or the black hole in the
 nonlinear interval  equations (18)-(20). But all these nonlinear
  equations  lead to Schwarzschild's time dilation  $d\tau/dt
 \approx 1 - GMr^{-1}$ [2]
under weak gravitational fields $GMr^{-1}\ll 1$ and non-relativistic
velocities $(1 - \beta) \ll 1$.

The following summary of  main relations between the  proper
metric tensor, $g_{\mu\nu} \equiv g^{_N}_{\mu\nu}(x)$, the proper 
four-velocity, $V_{\mu}\equiv m_{_N}^{-1}P_{_N\mu}(x)$, and the field
potential,
$U_\mu \equiv U_{_N\mu} (x) \equiv B_\mu^{\neq_N}(x)+S^{_N}_\mu(x)$,
may be useful
for our references,            
\begin{eqnarray}
  {g}_{oo} = (1 +  \beta U_o )^2,    
    \   {g}_{oi} = (1 +   \beta U_o )
 \beta U_i, \
 { g}_{ij} = {\beta^2 }U_iU_j   - \delta_{ij}, 
\nonumber \\
{ g}^i = - { g}^{oi} = \gamma^{ij} { g}_j = { g}_i = - { g}_{oi}
{ g}_{oo}^{-1} =
  -\beta U_i(1 + \beta U_o )^{-1},
\nonumber \\
 g^{oo} = g^{-1}_{oo} -
 g_i g^i
= (1 - \beta^2 U_iU_j \delta^{ij})(1 + \beta U_o)^{-2}, \ 
\gamma_{ij} = \gamma^{ij}=  -g^{ij} = \delta_{ij}, 
\nonumber \\
V_\mu = \{\beta^{-1} + U_o; -\beta^{-1}v_i+U_i \}={\cal V}_\mu+
U_\mu=g_{\mu\nu}V^\nu ,
\nonumber \\  
V^\mu =\{  (\beta^{-1} + U^o) \ ; \ V^i  \} =   \{\beta^{-1} - 
(U_o + U_i v^i )(1 + \beta U_o)^{-1}\  ;\ \beta^{-1} v^i \},
\nonumber \\
 \ V_\mu V^\mu = g_{oo} 
(V^o - g_i V^i)^2 - \delta_{ij}
 V^iV^j = V_o^2g^{-1}_{oo}
- \beta^{-2} v^2    =  1;
\nonumber \\
Gauge: S^o\equiv [S_o(1-\beta B^{\neq_N}_iv^i)(1+\beta B^{\neq_N}_o)^{-1}
 + S_iv^i](1+\beta U_o)^{-1} = 0 ,
\nonumber \\  
with \ U^o\equiv B_{\neq_N}^o + S^o = B_{\neq_N}^o = (B^{\neq_N}_o +\beta
 B^{\neq_N}_iv^i)(1+\beta B^{\neq_N}_o)^{-1},
 \nonumber \\ S^i\equiv 0,\ (1-\beta U_iv^i)(1+\beta U_o)^{-1} = (1-\beta
  B^{\neq_N}_iv^i)(1+\beta B^{\neq_N}_o)^{-1},
\nonumber \\
V^\mu  =  \{\beta^{-1} - (B^{\neq_N}_o + B^{\neq_N}_i v^i )(1 + \beta
B^{\neq_N}_o)^{-1}\  ;\ \beta^{-1} v^i \},
\nonumber \\
S_\mu V^\mu=0, \ S_\mu S^\mu=0, \ V_\mu V^\mu = ({\cal V}_\mu + 
B_\mu^{\neq_N} + S_{\mu}) V^\mu = 1. 
\end{eqnarray}

It is interesting that the contravariant component 
$P_{_N}^i \equiv m_{_N}V^i$ = $m_{_N}\beta^{-1} v^i$ does
 not depend on external potentials (because $U^i \equiv 0$), while
 the covariant three-momentum $P_{_N i }= - m_{_N}\beta^{-1}
 v_i + m_{_N}U_i$ depends on them. 
This means that the electromagnetic potentials $A_\mu $ and
 $A^\mu$, as well as the total potentials $U_\mu$ and $U^\mu$, 
are not themselves four-vectors.

The metric tensors $g_{_N}^{\mu\nu}$ can not be applied for rising
indexes of any one summand in  $m_{_N}V_\mu$ = $m_{_N}{\cal V}_\mu +
m_{_N}B_\mu + m_{_N}\partial_\mu \phi_{_N} + q_{_N}A_\mu$, despite
$g^{\mu\nu}_{_N}V_\mu = V^\nu $ for the total (canonical) four-velocity
  $V_\mu \equiv V_{_N\mu}$ of the charged mass N.
The physical gauge $S_\mu  P^\mu = 0$ in $V_\mu P^\mu =
({\cal V}_\mu + B^{\neq_N}_\mu + S_\mu )P^\mu = m_{_N}$ means that the
self-potential (and the charge $q_{_N}$) does not directly contribute
to the particle rest mass $m_{_N}$.

Their is no metrics to relate the self-potentials $S_\mu\neq 0$ and
$S^\mu = 0$, which are associated, in particular,  with the  electromagnetic
 potentials
$q_{_N}A_\mu$ and $q_{_N}A^{\mu}$.
A massless electric charge, $q_{_N} \neq 0$ with $m_{_N} = 0$, can not exist
 in practice because the electromagnetic part, $q_{_N}A_{\mu}(x)$, of the
 canonical four-momentum  does not  maintain a definite tensor nature in
 the proper four-space $x_{_N}^\mu$.

\bigskip \bigskip
\noindent {\bf 5. Point macroscopic and extended microscopic particle states}

\bigskip
\noindent {\it 5.1. The macroscopic geodesic motion of the point charge}
\bigskip

The coupled gravitational and electromagnetic potentials  in (2) or (14)
 can lead in principle to balanced vector forces under the free motion of the
 charged particle.
That is why we may generalize the equivalence principle  
on the proper canonical four-space $x_{_N}^\mu$ with the Christoffel
connections (17).
A macroscopic geodesic equation  may be derived due to the covariant
 conservation
of the localized particle four-momentum $P_{_N\mu}(\xi_{_N})$  at every
 path
 point $\xi_{_N}$ in the curved canonical four-space, where
$V_{_N\mu}V^\mu_{_N} \equiv 1$,  $V_{_N}^\nu \nabla^{_N}_\mu V_{_N\nu}
\equiv 0$, and $\Gamma^\mu_{\rho\lambda} = \Gamma^\mu_{\lambda\rho}$,
\begin {eqnarray}
{{D P_{_N\mu} (\xi_{_N})}\over ds_{_N}(\xi_{_N})}={{V_{_N}^\nu }} 
(\nabla_\nu P_{_N\mu} - \nabla^{_N}_\mu P_{_N\nu})_{x\rightarrow \xi_{_N}}
 = {{P_{_N}^\nu}}  (\partial_\nu
{ V}_\mu -\partial_\mu { V}_\nu)
\nonumber \\=P^\nu_{_N}(\partial_\nu
{\cal V}_\mu -\partial_\mu{\cal V}_\nu)
  - P^\nu_{_N}(\partial_\mu B_{\nu}
- \partial_\nu B_{\mu})  -
q_{_N}V^\nu_{_N}(\partial_\mu A_{\nu}
- \partial_\nu A_{\mu})
\nonumber \\
 = m_{_N} V^\nu (\partial_\nu {\cal V}_\mu -\partial_\mu{\cal
  V}_\nu)_{x\rightarrow \xi_{_N}}
  - m_{_N}V^\nu \sum_{_K}^{_K\neq_N} (-Gm_{_K})f_{\mu\nu}^{_K-}(\xi_{_N})
\nonumber \\
-q_{_N}V^\nu \sum_{_K}^{_K\neq_N} q_{_K}f_{\mu\nu}^{_K+}(\xi_{_N})
-(q^2_{_N}-Gm^2_{_N})V^\nu f^{_N}_{\mu\nu}(x)_{x\rightarrow \xi_{_N}}  = 0. 
\end {eqnarray}

The retarded Lorentz force for the electric charge, $q_{_N}V^\nu_{_N}
(\partial_\mu A^{\neq_N}_{\nu}
- \partial_\nu A^{\neq_N}_{\mu})$ = $q_{_N}V^\nu_{_N}\sum_{_K}^{_K\neq_N}
 q_{_K}(\partial_\mu a^{_K+}_{\nu}
- \partial_\nu a^{_K+}_{\mu})$ = $q_{_N}V^\nu_{_N}\sum_{_K}^{_K\neq_N}
q_{_K}f_{\mu\nu}^{_K+}(\xi_{_N})$,
is accompanied in (22) by the advanced gravitational force
for the mass with the similar vector nature, $ m_{_N}V^\nu_{_N}
(\partial_\mu B^{\neq_N}_{\nu}
- \partial_\nu B^{\neq_N}_{\mu})$=$m_{_N}V^\nu_{_N}\sum_{_K}^{_K\neq_N}
(-Gm_{_K})(\partial_\mu a^{_K-}_{\nu}
- \partial_\nu a^{_K-}_{\mu})$ = $m_{_N}V^\nu_{_N}\sum_{_K}^{_K\neq_N}
(-Gm_{_K})f_{\mu\nu}^{_K-}
(\xi_{_N})$.
The self-force $(q^2_{_N}-Gm^2_{_N})V^\nu f^{_N}_{\mu\nu}(x)_{x\rightarrow
\xi_{_N}}$ originates from the net energy loss under generation of
electromagnetic and gravitational waves. Below this self-force will be
bound with the integral changes of  the self-field tensor density
$\theta^\mu_\nu (x)$ of the microscopic particle states  over all vacuum
 points $x\neq\xi_{_N}$. Notice for (22) that the gauge term is absent
 due to a tensor equality  $ P^\nu (\partial_\nu \partial_\mu \phi_{_N}
  -\partial_\mu \partial_\nu \phi_{_N}) = 0 $
and there are only three
independent equations  due to scalar equalities
 $P^\mu D P_\mu \equiv 0$ and $P^\mu P^\nu
(\partial_\nu V_\mu -\partial_\mu V_\nu)\equiv 0$. 

The macroscopic geodesic equations (22), based on the canonical four-space
with metric relations (15)-(17) and the physical gauge 
$S^o = S^\mu = 0$,  $({\cal V}_\mu + B^{\neq_N}_\mu)V^\mu=1$,
corresponds to the classical Minkowski-Lorentz analogue
 $\!Du_\mu/ds$
$=$ $m^{-1}_{_N}q_{_N}u^\nu (\partial_\mu A_\nu - \partial_\nu A_\mu) $,
 based on the four-space concept with
pure mechanical connections and  $u_\mu u^\mu = 1$.     

One can  read the general geodesic equation (22) in the non-relativistic
 limit, where
$q_{_N}(E + v\times B)_i + m_{_N}(E_g + v\times B_g)_i$ = 
$m_{_N}[  V^o(\partial_i {\cal V}_o - \partial_o {\cal V}_i) +  V^j
(\partial_i {\cal V}_j - \partial_j {\cal V}_i)
 ]  $ $\approx m_{_N} [\partial_o {v}_i + (v^j\nabla_j)v_i ] = m_{_N}
 dv_i/dt$
and $ v^i (m_{_N}E_{gi} + q_{_N}E_i) = m_{_N}v^i dv_i / dt$.
A self-action here may be found from a non-divergence part of the proper
four-potential [8], $a_{_N\mu}(x)_{x \rightarrow \xi_{_N}}\approx
\{0, -2{\dot v}/3 \}$, which lead to $E^{self}_i = q_{_N}2{\ddot v}/3$,
$E^{self}_{gi} = -Gm_{_N}2{\ddot v}/3$, $B_i=B_{gi}=0$, and $m_{_N}{\dot v}_i
= - 2(Gm^2_{_N} - q^2_{_N}){\ddot v}_i/3$.   

The physical gauge restriction, $S^{_N}_{\mu}(x)P^\mu_{_N}(x)=0$, and
 the covariant
conservation $DP^\mu_{_N}/ds_{_N} = 0$, lead to a spin-type relation
for the self-momentum $m_{_N}S_{_N\mu}$,
\begin {equation}
P_{_N}^\mu{{DS_{_N\mu}}\over ds_{_N}} = m_{_N}V^\mu \left (
{{dS_{_N\mu}}\over ds_{_N}}  - \Gamma^\rho_{\mu\nu}V^\nu S_{_N\rho} \right ). 
\end {equation}
This equation
reads $V^oV^o\Gamma^i_{oo}S_{i} = 0$ in the particle rest frame, where
 $V^i = 0$ and $S_{o} = 0$. From here one finds a scalar conservation
 $\partial_o (m^2_{_N}S_{i}S_{j}\delta^{ij}) = 0$ with the proper
 self-momentum $m_{_N}S_{i}(\xi_{_N})=m_{_N}\partial_i \phi_{_N}(\xi_{_N})
  + (q_{_N}^2-Gm^2_{_N})a_{_Ni}(\xi_{_N})$ $\neq 0$, while $m_{_N}S_{o}
  (\xi_{_N})= m_{_N}\partial_o \phi_{_N}(\xi_{_N}) -
(Gm^2_{_N}-q_{_N}^2)a_{_No}(\xi_{_N}) = 0$, in the absence of external
gravitational and electromagnetic fields. The self-potential $S_{_N\mu}$
has only three degrees of freedom and it is not a four-vector.
This potential may be associated with an internal angular momentum 
or spin of the elementary object N with the net rest mass $m_{_N}$.

\bigskip {\it 5.2.  Maxwell-type vector equations for the particle mass
and charge}

\bigskip
The macroscopic geodesic motion of the point source-outlet N along the
classical path $\xi_{_N}$ depends on   elementary electromagnetic,
$q_{_N}f_{\mu\nu}^{_K+}(\xi_{_N})$,
and gravitational, $-Gm_{_K}f_{\mu\nu}^{_K-}(\xi_{_N})$, fields related
by zero four-intervals with distant sources-outlets K at point vertices
 $\xi_{_K}$.  The equation (22) is
 not complete if we do not know how to relate these skew-tensor fields
  with
their moving sources-outlets.

 One way to relate the gravitational field to the mass corresponds to
 the tensor Einstein equation [1], which may be derived, for example,
  after the Hilbert variation [17] of a system action with respect to
   ten  variations $\delta g_{\mu\nu}$  ($g_{\mu\nu}V^\mu V^\nu$ $\neq$
    1 before the variations). Recall that
there is no one ten-component field in (15)
behind the symmetric tensor $g^{_N}_{\mu\nu}$  and its ten degrees of
freedom
have different physical origin. These degrees of freedom may be also 
traced in the four-vector variable $P_{_N\mu}(x)$. We have first to define 
an elementary action for a particle N and its self field in order to
select dynamical variables and derive Lagrange variational equations.  

At first we accept an introduction of the particle in terms of extended
 microscopic states over all vacuum space by keeping the localized or
 point particle state  exclusively for space-time averaged, macroscopic
  relations. The wave-particle dualism, propounded in 1905 by Einstein
  in his quantum theory of light, was fruitfully inverted by de Broglie
   in 1923, when he first departed from the point particle notion and
developed the field analogy of the particle with the light wave [18].
By following this verified way, we may postulate that the particle is
 not localized under its microscopic states no matter the classical or
  the wave approach is taken into consideration. In other words, we are
   going to reinvest the Einstein - de Broglie particle-wave dualism from
   Wave Mechanics back to
Classical Electrodynamics and Gravitation through a postulated concept
of the dual particle states (point macroscopic and nonlocal microscopic
ones). The observed localization of the classical particle in our approach
will take place only for the averaged, macroscopic state.

The particle integration into the classical field
 structure had been anticipated by Einstein at the end of his life:
 "We could regard matter as being made up of regions
of space in which the field is extremely intense... There would be no
room in this new physics for both field and matter, for the field would
 be the only reality" (translation [19]). His analysis of the Maxwell-Lorentz
  equations revealed the following: "The combination of the idea of a
  continuous fields with that of discontinuous material points in space
   strikes me as contradictory. A coherent field theory would consist
   exclusively of continuous elements not only in time but in every
   point of space. Hence the material particle cannot possible be a
    fundamental concept in any field theory. Thus, quite apart from
     the fact that it ignores gravitation, Maxwell's theory cannot be
      considered as a complete theory"(translation [20]). The Lorentz
       introduction of point field sources into microscopic equations
        is "an attempt which we have called intellectually unsatisfying" [20],
        and dissatisfied Einstein criticised his gravitational equation with
        the point sources: "it resembles a building with one wing build of
        resplendent marble, and the other build of cheap wood". The problem
         of point particles in the classical field equations has not been
         resolved satisfactorily yet that may be considered as a motivation
          for our introduction of the extended microscopic states along with
           the averaged macroscopic state for the same entity, called the
           particle.

The Dirac delta-operator, with $\int dx^4{{P_{_N\mu} (x) {\delta}^3 ({\bf x}
-{\mbox {\boldmath $\xi $}}_{_N}[p]) {\delta}(x^o-\xi^o_{_N}[p])}}$ $=$
$P_{_N\mu} (\xi_{_N}[p])$, will assist us to interpret the particle
in terms of the extended microscopic states (or continuous virtual
 fluctuations) with the local particle four-momentum $P_{_N\mu}(x)=
 m_{_N}V_\mu(x)$ and the
local four-velocity $V^\mu(x)$ at all vacuum points $x^\mu\neq \xi^\mu_{_N}$. 
We will consider the particle N in microscopic relations as a nonlocal
 continuous fraction of an elementary object N, where the relevant particle
  densities are locally bound with induced self fields $q_{_N}f_{\mu\nu}^{_N}
  (x)_{x\neq
\xi_{_N}}$ and $-Gm_{_N}f_{\mu\nu}^{_N}(x)_{x\neq\xi_{_N}}$. Such
interpretation of the particle for microscopic states can be agreed both
 with macroscopic particle peculiarities and with an introduction of the
  material vacuum through a global intersection of elementary particle-field
  objects with zero elementary mass/charge/energy densities. Again, the
  particle entity without its self-fields is not a complete elementary
  object in our consideration.

Now we can define the tensor density $W^{_N}_{\mu\nu}(x)$ $\equiv$
$\nabla^{_N}_\mu {P}_{_N\nu}(x)$ - $\!\nabla^{_N}_\nu P_{_N\mu}
(x) $ = $ \partial_\mu {P}_{_N\nu}(x)$ $-$ $\partial_\nu P_{_N\mu}
(x)$ of the extended particle states and the self-field tensor   
$f^{_N}_{\mu\nu}(x) \equiv$ $\nabla^{_N}_\mu
{a}_{_N\nu}(x)$ - $\!\nabla^{_N}_\nu a_{_N\mu}
(x) $ =$\partial_\mu {a}_{_N\nu}(x)\!-\!\partial_\nu a_{_N\mu}
(x)$ at the same points of the proper four-space $x_{_N}^\mu=x^\mu$
(for brief). A local coupling of the microscopic particle states and
their fields
takes place overall $x^\mu_{_N}$ and a scalar contraction of their
tensor densities should be put into the
 microscopic action ${\cal S}^{mic}_{_N}$ of the complete elementary
  particle-field object N,

 \begin {eqnarray}
{\cal S}^{mic}_{_N}\!=\!-\!\int\!{d^4x}{\sqrt{-g_{_N} } }  
 \int d\xi^o_{_N}[p]
{{\delta^3({\bf x}\!-\!{\mbox{\boldmath $\xi $}}_{_N}[p])
\delta(x^o\!-\!\xi^o_{_N}[p])\!P^{x\neq \xi_{_K}}_{_N\mu}(x)}
\over {\sqrt {-g_{_N}(x)} } }{{dx^\mu}\over d\xi^o_{_N}[p]}
\nonumber \\
-\int_{x\neq \xi_{_K},\xi_{_N}}
{{d^4x{\sqrt {-g_{_N}(x)}} }\over 16\pi }  
\left [{{{  g_{_N}^{\mu\rho}(x)
g_{_N}^{\nu\lambda}(x) W^{_N}_{\mu\nu}(x)
        f^{_N}_{\rho\lambda}(x) }} }  +
 g^{_N}_{\mu\nu}(x)G^{-1}r^{\mu\nu}_{_N}\right ]\!
\nonumber \\
=\!-\!\int\!{d^4x}{\sqrt {-g_{_N}}}    
\left (\!P^{x\neq \xi_{_K}}_{_N\mu} 
{{{\delta}^3({\bf x}\!-\!{\mbox {\boldmath $\xi $}}_{_N}[x^o])\!dx^\mu}
\over {\sqrt {-g_{_N}(x)}}dx^o}
\!+\!{{W^{x\neq \xi_{_K}}_{_N\mu\nu}
f_{x\neq\xi_{_N}}^{_N\mu\nu}}\over 16\pi} + 
{{r_{_N}(x)} \over 16\pi G}\right ),
\end {eqnarray}
where ${\sqrt {-g_{_N}(x)}}\equiv 
{\sqrt {\gamma_{_N} g^{_N}_{oo}(x)}}\equiv 
{\sqrt {g^{_N}_{oo}(x)}}$, 
$r_{_N}(x) \equiv g^{_N}_{\mu\nu}
(x)r_{_N}^{\mu\nu}(x)_{x\neq \xi_{_N},\xi_{_K}}$ is the scalar curvature of 
the four-space $x^\mu_{_N}$ at peculiarity free points,
$f_{_N\mu\nu}(x)$ is not define at $\xi_{_N}$,
 and $P_{_N\mu}(x)$ is not define at all $\xi_{_K}$, 
when $\xi_{_K}\neq \xi_{_N}$.

 By applying the delta-operator in (24), one may
find an action ${\cal S}^{mac}_{_N}$ =  $-\int dp P_{_N\mu}(\xi_{_N}
[p])(d\xi_{_N}^\mu[p]/dp) $ for the
macroscopic, localized particle at its classical path $\xi_{_N}$, where
there are no self-fields in our definition. The path variations $\delta
 \xi_{_N}[p]$ at these bare particle points in  ${\cal S}^{mac}_{_N}$
 lead to  $m_{_N}V^\nu_{_N}(\xi_{_N})W^{_N}_{\mu\nu}(\xi_{_N})$ = 0,
 {\it i.e.} to the macroscopic geodesic equation (22) under its first
  integral $V_{_N\mu}(\xi_{_N}) V^\mu_{_N}(\xi_{_N}) = 1$. The geodesic
   equation for the extended microscopic states at the field points $x\neq
    \xi_{_N}$ will be derived below after the
action variations  with respect to $\delta x_{_N}^\mu$.

The variations of (24) with respect to  $\delta P_{_N\mu}(x)$ at pure
 field points $x\neq \xi_{_N}, \xi_{_K}$ correspond to a microscopic
 Lagrange equation
\begin {equation}
 I_{_N}^\nu(x)_{x\neq\xi_{_N},\xi_{_K}} \equiv  i_{_N}^\nu(x) - 
{{ \nabla^{_N}_\mu f_{_N}^{\mu\nu}(x)}\over 4\pi }
\equiv {{{\delta}^3 ({\bf x}-{\mbox {\boldmath $\xi $}}_{_N}[x^o]) dx^\mu }
\over {\sqrt {-g_{_N} (x) } } dx^o }  
 - {{ \nabla^{_N}_\mu f_{_N}^{\mu\nu}(x)}\over 4\pi } = 0,
\end {equation}
where $f_{_N}^{\mu\nu}(x) \equiv g_{_N}^{\mu\rho}(x)
g_{_N}^{\nu\lambda}(x)$
${f}^{_N}_{\rho\lambda}(x)_{x\neq
\xi_{_N},\xi_{_K}}$ and ${\sqrt {-g_{_N}(x)}}\equiv {\sqrt {g^{_N}_{oo}(x)}}$.
Notice that not all components of the skew-symmetric
tensors are independent under variations [19]:
 the relations $\delta W^{_N}_{\mu\nu}(x)$ =
$ - \delta W^{_N}_{\nu\mu}(x)$ were taken into account. 
The variable $P_{_N\mu}(x)$ has four degrees of freedom before
the variations because $P_{_N\mu}dx^\mu/dp \neq const $ in the action (24). 
We may define $dp = ds_{_N}$ for the path parameter $p$ after the
 variations and then use $V_{_N\mu}(x)V^\mu_{_N}(x)=1$ in the equations
  of motion.

At first glance the Maxwell-type variational equation (25) seems not new
 at all. The only point it is not specified neither for the mass, nor for
  the electric or other kind of charge.
This basic equation for the microscopic particle states and their fields
manifests in general that any kind of the particle matter four-flow
$i_{_N}^\nu(x)$ or pre-current is to be locally screened by an induced
 appropriate pre-field $f_{_N}^{\mu\nu}(x)$ at all peculiarity free points,
  called vacuum or field points in macroscopic physics. Indeed, different
  solutions of this microscopic equation may be associated with differently
   charged particle-field matter. Below we apply (25) only to the mass and
   to the electric charge in order to derive macroscopic equations and to
   relate
the macroscopic gravitational fields to the advanced Lienard-Wiechert
potentials, while the macroscopic electromagnetic fields with the retarded
 ones.  But first we rewrite (25) for the relevant microscopic
equations with the active gravitational mass $m_{_N}$ 
and the electric charge $q_{_N}$, respectively,     
\begin {equation}
{\cases 
{ -4\pi G {\sqrt {g^{_N}_{oo}}}m_{_N}i_{_N}^\nu(x)_{x\neq\xi_{_N},\xi_{_K}}
=  \partial_\mu  {\sqrt {g^{_N}_{oo}}}[-Gm_{_N}f_{_N}^{\mu\nu}(x)]_{x\neq
\xi_{_N},\xi_{_K}}\cr  
\ \ \ 4\pi {\sqrt {g^{_N}_{oo}}}q_{_N}i_{_N}^\nu(x)_{x\neq\xi_{_N},\xi_{_K}}\
=\  \partial_\mu {\sqrt {g^{_N}_{oo}}}q_{_N}f_{_N}^{\mu\nu}(x)_{x\neq
\xi_{_N},\xi_{_K}}\cr
}}.
\end {equation}

Here there is only a local coupling of the fields with the microscopic
currents
$\{-Gm_{_N}/q_{_N}\}i^\mu_{_N}(x)_{x\neq \xi} $ $\equiv$ 
$\{-Gm_{_N}/q_{_N}\} {{{\delta}^3 ({\bf x}-{\mbox {\boldmath $\xi $}}_{_N}
[p]) dx^\mu } / {\sqrt {-g_{_N} } } dx^o }$ of the extended particle in
$x^\mu\neq \xi^\mu_{_N}$. The microscopic particle currents (or virtual
 particle fluctuations) reshaping microscopic fields overall space with
an infinite speed $C=\infty$, otherwise another Dirac operator, ${\hat
\delta}^4(x-\xi_{_N}[p])= {\delta}^3 ({\bf x}-{\mbox {\boldmath $\xi
$}}_{_N}[p])\delta (x^o - \xi^o_{_N}[p] \pm C^{-1}|{\bf x}-{\mbox
{\boldmath $\xi $}}_{_N}[p]|)$, is to be used in the action (24). The
 nonlocal nature
of the particle on the microscopic level is responsible for instantaneous
updating of both outgoing and incoming microscopic fields that is not in
disagreement with the Special Relativity restrictions for macroscopic 
fields.

The advanced/retarded relations of macroscopic self-potentials with their
mass/charge at $\xi_{_N}$ do arise from (26), but only for
the macroscopic states averaged over all microscopic states (or virtual
fluctuations for macro-world) under "small macroscopic - large microscopic"
time scales $\Delta x^o \sim t^o$. At these scales the particle and its
fields disintegrate
in space, and the macroscopic generalization of (25) takes the following
form,
$$
{\cases {i_{_Nmac}^\mu (x)_{x\neq\xi_{_N}} = 0, \ \ f^{\mu\nu}_{_Nmac}
(x)_{x\neq\xi_{_N}} \neq 0, \ \   \nabla^{_N}_\mu f^{\mu\nu}_{_Nmac}
(x)_{x\neq\xi_{_N}} = 0, \cr
 i_{_Nmac}^\mu (\xi_{_N}) \neq 0, \ \  f^{\mu\nu}_{_Nmac}(\xi_{_N}) = 0, \ \ 
\nabla^{_N}_\mu f^{\mu\nu}_{_Nmac}
(x)_{x = \xi_{_N}} = 0.
  }   } \eqno (25 a)
$$

 The macroscopic disintegration of masses or 
charges with their advanced or retarded, respectively, fields in (25a)
should be considered as a fundamental property of this kind of matter.
In principle, one may find some more field solutions of
the general equation (25), which are not related to space disintegration 
with extended charges. Such charges cannot be observed in macro-world
and they are out of plans to consider macroscopic gravity and electrodynamics.

One may say that the microscopic particle states or virtual particle
fluctuations
are screened completely by locally induced charged fields, and every
elementary material object has neither the net mass  four-current density,
$m_{_N}{I}_{_N}^\nu(x)_{x\neq
\xi} = 0$, nor the electric four-current
density, $q_{_N}{I}_{_N}^\nu(x)_{x\neq
\xi} = 0$, at all field points. There are no screening self-fields in the
vertex $\xi_{_N}$, where the elementary object is represented by only the
 particle fraction. Thus the particle exhibits in $\xi_{_N}$ its finite
 mass/charge four-current on behalf of the complete elementary object,
 which becomes available for observations at this point.

Another microscopic  equations and their macroscopic generalizations for
the incoming and outgoing field densities may be derived directly from
their skew-symmetrical tensor structures $f^{_N}_{\mu\nu}(x)_{x\neq
\xi_{_N}}$,
\begin {equation}
{\cases{
 (-Gm_{_N})[\partial_\mu f_{_N\nu\delta}(x)
+\partial_\nu f_{_N\delta\mu}(x)
+ \partial_\delta f_{_N\mu\nu}(x)
]^{mic/mac}_{x\neq \xi_{_N}}
 = 0 \cr
\ q_{_N}[\partial_\mu f_{_N\nu\delta}(x)
+\partial_\nu f_{_N\delta\mu}(x)
+ \partial_\delta f_{_N\mu\nu}(x)
]^{mic/mac}_{x\neq \xi_{_N}}
 = 0 \cr
}.}
\end {equation}

Notice that the macroscopic equations (25a) and (27) admit retarded wave
solutions for the field $q_{_N}f_{_Nw}^{\mu\nu}(x)_{x\neq \xi_{_N},\xi_{_K}}$
 with a zero electric current,
$q_{_N}\nabla_\mu f_{_Nw}^{\mu\nu} $ $\equiv$ 
$q_{_N}{ {g^{-1/2}_{oo}}}\partial_\mu 
{\sqrt {g_{oo}}} f_{_Nw}^{\mu\nu}$ = 0, and advanced wave solutions
 $[-Gm_{_N}f_{_Nw}^{\mu\nu}(x)]_{x\neq \xi_{_N},\xi_{_K}}$ with a zero
  mass current,
$(-Gm_{_N})\nabla_\mu f_{_Nw}^{\mu\nu}$ $\equiv$ $-Gm_{_N}{g^{-1/2}_{oo}}
\partial_\mu {\sqrt {g_{oo}}}
 f_{_Nw}^{\mu\nu}$ = 0. They are anti-waves to each other because
 the spherical electromagnetic wave is moving from its point source
  $\xi_{_N}$, while the spherical gravitational wave is moving with the same
   light speed $c$ toward its point outlet $\xi_{_N}$.

 Only the particle four-currents may induce the interaction
 fields based on the uncharged pre-field  $f_{\mu\nu}$. These fields
 affect another particles and couple material entities with nonzero
 mass/charge four-currents.
 The gravitational/electromag\-ne\-tic mass/charge dimensional waves 
from (26)-(27) have formally the same mass / charge density, respectively,  
as their parent particle.
But these "massive" and "charged" fields with zero mass and charge
four-currents  do not induce any interaction fields. In other words,
 the
gravitons and the photons have only zero active gravitational masses
(as well as zero passive and inertial masses, and zero passive and active 
electric charges) and they do not couple to each other in our consideration.

It is important for macroscopic causality that the advanced gravitational 
wave or their packet would come from infinity to the point vertex-outlet 
only through infinite time intervals and, therefore, this gravitational
wave never be registered in practice or cross the retarded electromagnetic
 wave. From the other side, the instantaneous generation of gravitational
 waves in infinity takes energy from the accelerated mass and lead, for
 example, to the radiation damping in oscillation of binary pulsars. In
  general the mass exhibits inertia or resists to any acceleration by
  generating the advanced waves in infinity, while the non-inertial
  charge welcome self-accelerations by generating the retarded waves.

We should recall  from [5], for example, for the uniformly moving particle
that  its Newton and Coulomb forces, derived from the advanced and
retarded Lienard-Wiechert potentials, respectively, are both directed
toward the instantaneous particle position, rather than toward different
(advanced and retarded) positions. These forces are always parallel due
to the Lorentz invariance in all inertial frames of references, including
 the particle rest frame, and forces cannot be promptly used for verifying
  the advanced nature
of macroscopic gravitation. Aberrations of Sun's electromagnetic and
gravitational waves would have different signs, could causality be violated
 and the gravitation wave be registered one day.

\bigskip{\it 5.3.  Superfluid behaviour of microscopic particle states }

\bigskip

Now we derive one more Lagrangian equation by varying the global ensemble
action $\sum^{\infty}_{_N}{\cal S}_{_N}^{mic}$ with respect 
to  $\delta a_{_N\mu}$,
\begin {equation}
\nabla_\mu W^{\mu\nu}_{_N} = 0,
\end {equation}
where we used $\delta P_{_K\mu} (a_{_N\mu}) = const \delta a_{_N\mu} $
 and (25).

We also may write directly from the  definition   
 $W^{_N}_{\mu\nu}(x)$
$\equiv$ $  \nabla^{_N}_\mu
{P}_{_N\nu}(x)-\nabla^{_N}_\nu P_{_N\mu}(x)$ another relation
 for the microscopic particle states in all field points $x\neq \xi_{_N},
  \xi_{_K}$ or in vacuum,
\begin {equation}
\partial_\mu W^{_N}_{\nu\delta}(x)
+\partial_\nu W^{_N}_{\delta\mu}(x)
+ \partial_\delta W^{_N}_{\mu\nu}(x) = 0
\end {equation}  with  
$m_{_N}[ \partial_\mu {M}^{_N}_{\nu\delta}(x)
+ \partial_\nu {M}^{_N}_{\delta\mu}(x)
+ \partial_\delta {M}^{_N}_{\mu\nu}(x)] = 0$ when $q_{_N} = 0$.

One could formally decompose  the total particle  four-momentum
$P^{_N}_{\mu}
(x)$ in (2) into a gravitomechanical
 part and an electrical one.
Then the  tensor density $W^{_N}_{\mu\nu}(x)$
 of extended particle states in vacuum would be formally divided
into  a gam\-vi\-to\-mechanical part  and
an electric part (with  $q_{_N}$),
$ W_{_N\mu\nu} (x)$ = $ 
m_{_N} M_{_N\mu\nu}(x) +
q_{_N}  F_{\mu\nu}(x)$, where
$M_{\mu\nu}(x)$  $\equiv $
${w}_{\mu\nu}(x)  + G_{\mu\nu}(x) $,
${w}_{\mu\nu}(x) \equiv  \partial_\mu {\cal V}_{\nu} (x)
- \partial_\nu {\cal V }_{\mu}(x)$,
${\cal V}_\mu(x) \equiv \{\beta^{-1}, -\beta^{-1}v_i \}$, $\beta
= {\sqrt {1 - v_i v^i}}$,
$G_{\mu\nu} (x)\equiv  \partial_\mu B_{\nu} (x)
- \partial_\nu B_{\mu}(x)$
$ =\sum_{_K}^{all} m_{_K}$ $ [\partial_\mu {a^{-}_{_K\nu}} (x)
 - \partial_\nu {a^{-}_{_K\mu}}]_{x\neq \xi_{_K}}$,
and  $F_{\mu\nu}(x) $ $
  \equiv $ $\partial_\mu A_\nu (x)$ 
- $\partial_\nu A_\mu (x)$
$ =\sum_{_K}^{all} q_{_K}$ $ [\partial_\mu {a^{+}_{_K\nu}}(x)
 - \partial_\nu {a^{+}_{_K\nu}}(x)]_{x\neq \xi_{_K}}$.

The vortexless partial solutions, $W_{_N}^{\mu\nu}(x) \equiv
g_{_N}^{\mu\rho}g_{_N}^{\mu\lambda}
W^{_N}_{\rho\lambda} = 0$ and $W^{_N}_{\mu\nu}(x) \equiv \partial_\mu
 P_{_N\nu}(x)-\partial_\nu P_{_N\mu} (x)= 0$, of the general equations
  (28)-(29) can be written through a gradient,
 $\partial_\mu \Upsilon_{_N}(x)$,
of the single valued scalar potential $\Upsilon_{_N}(x)$, with
$\partial_\mu \partial_\nu \Upsilon_{_N}(x) =
\partial_\nu \partial_\mu \Upsilon_{_N}(x)$ and
\begin {eqnarray}
m_{_N}V_{_N\mu}(x)\!=\!m_{_N}{\cal V}_\mu\!+\!m_{_N}
B_\mu(x)\!+\!q_{_N}A_\mu(x)\!+\!m_{_N}\partial_\mu\phi_{_N}(x)=
 m_{_N}\partial_\mu
\Upsilon_{_N}(x).
\end {eqnarray}

One may associate this scalar $\Upsilon_{_N}(x) = \int V_{_N\mu}dx^\mu
 + const= s_{_N}(x)+const$ with the proper four-interval $s_{_N}$, because
 $d\Upsilon_{_N}$
= $dx^\mu \partial_\mu \Upsilon_{_N} $ = $dx^\mu V_{_N\mu} $ = $ds_{_N}$.
The potential microscopic states (30)
correspond to the well known 
London supercurrent of Cooper's carriers in electromagnetic fields. For
this reason the microscopic particle states  may be considered as
 superfluid virtual fluctuations of the particle in its proper four-space.
   Vortexless motion of microscopic particle densities in vacuum can be
    read through the following three-vector functions
                     \begin{eqnarray}
{g}_{i}(x)\equiv \partial_o {\cal V}_i -\partial_i {\cal V}_o \equiv
-{\partial_o} v_i \beta^{-1}
- \partial _i  \beta^{-1}
\nonumber \\ = - {G}_{i}(x) - m_{_N}^{-1}q_{_N}
{ E}_{i}(x)
\equiv
{\partial_o}
 \beta^{-1} {\sqrt {g^{_N}_{oo}}}  g_i 
+ \partial _i
   \beta^{-1}( {\sqrt { g^{_N}_{oo}}} - 1 ),
\end{eqnarray}
\begin{eqnarray}
 h^i(x) 
 \equiv -{2^{-1}{e^{ijk}}} (\partial_j{\cal V}_k-\partial_k
{\cal V}_j)
\equiv \{ curl \beta^{-1}{\bf v} \}^i
\nonumber \\
= -  H^i(x) - m_{_N}^{-1}q_{_N}B_q^i(x)
\equiv -\{ curl \ \beta^{-1} {\sqrt{g^{_N}_{oo}}} 
{\bf g} \}^i,
\end{eqnarray}
where the universal tensor density $F_{\mu\nu}(x) $ forms
the  three-vector electric,
 ${{ E}_{i}(x) }$ $\equiv$ $F_{oi}(x)$,
 and magnetic, ${B}_{q}^i(x)$ $\equiv -e^{ijk}
F_{jk}(x)/2{\sqrt \gamma }$ ($e^{123}=1, {\sqrt \gamma }=1$), fields,
 acting locally on the electric charge $ q_{_N}$ of the extended vacuum
state of the particle N.
Gravity is implicated into  (31) - (32) through the universal tensor
  $G_{\mu\nu}(x)$ based on potentials $B_\mu(x)=-G\sum_{_K} m_{_K}
a^{-}_{_K\mu}(x)_{x\neq \xi_{_K}}  $,
with $ G_i(x)\equiv G_{oi}(x) \equiv \partial_o B_i - 
\partial_i B_o $ and ${ H}^i({ x})\equiv-2^{-1}{{e^{ijk}} }
  G_{jk}(x)\equiv-2^{-1}{{e^{ijk}}
 } (\partial_j B_k-\partial_k B_j)$.
We take that ${\sqrt \gamma} = 1$, $\beta = {\sqrt {1 - \delta_{ij} v^i
 v^j}}$,
$\{ curl \ {\bf a}\}^i$ $\equiv$ $(2 {\sqrt {\gamma}})^{-1}$
 $e^{ijk}(\partial _j a_k - \partial _k a_j)$
= $2^{-1}$
 $e^{ijk}(\partial _j a_k - \partial _k a_j)$, and
$div \ {\bf a}$ $\equiv$ $\gamma^{-1/2}\partial_i ({\sqrt \gamma} a^i)$
= $\partial_i a^i $ for flat three-space. 
Notice directly from (31)-(32) in the absence of external fields that
the  elementary gravitational and electromagnetic self-fields are locally
induced by the extended
particle velocity $v_i(x)$ at vacuum points $x\neq\xi_{_N}$.

The general equations (28)-(29) admit also non-potential microscopic 
states of the particle N with a finite vortex density, when 
$W^{_N}_{\mu\nu}(x)\neq 0$. A number of elementary vortices crossing a 2D
 surface $s^{\mu\nu}$,
$n\!=\!h^{-1}\!\int\!W^{_N}_{\mu\nu}ds^{\mu\nu}\!=\!h^{-1}
\!\int\!P^{_N}_{\mu}dx^\mu$, is to be an integer with respect to the Plank
constant introduced by Wave Mechanics.

A natural question arise: why one observes in practice only point, localized 
particles and do not observes directly their extended microscopic states or
  virtual particle fluctuations in vacuum points? In order to answer this
  conceptual question we should find the stress-energy tensor density of
  these microscopic states.

\bigskip
\noindent {\it 5.4. The stress-energy tensor density of the elementary
particle-field object}
\bigskip

 By varying the action (24) with respect to $\delta g^{_N}_{\mu\nu}(x)$ one
may  find the stress-energy tensor density of elementary matter  at all
 four-space points, including the  particle vertex $\xi_{_N}$.
The self-field is not induced at the vertex point, where the localized,
macroscopic  particle state  exhibits a nonzero energy-tensor,
$T_{_N}^{\mu\nu}(\xi_{_N}[s_{_N}])$ $\equiv$
 $m_{_N}V^\mu_{_N}(\xi_{_N}[s_{_N}])V^\nu_{_N}(\xi_{_N}[s_{_N}]) \neq 0
$. The localized particle is not screened by its self-field at the vertex
 point, where the bare particle is observed in practice. The macroscopic
 equation of free motion, $\nabla_\mu T_{_N\nu}^{\mu}(\xi_{_N})$ =
$ m_{_N} [V_\nu \nabla_\mu V^\mu(\xi_{_N})  + 
D V_\nu (\xi_{_N})]$ = 0, of the
localized particle is consistent
with its mass/charge four-currents 
conservation and with the macroscopic  geodesic
equation (22).

 In vacuum points, $x\neq\xi_{_N}, \xi_{_K}$, we fix  the four-vector
 $dx^\mu_{_N}$ in  $dx_{_N\mu}(x) = g^{_N}_{\mu\nu}(x)dx_{_N}^\nu $
and $V^{_N}_{\mu}(x)\!=\!dx_{_N\mu}(x) / 
[dx_{_N\nu}(x) dx_{_N}^\nu ]^{1/2}$ under the variations
$\delta g^{_N}_{\mu\nu}(x)$.
Then the  variation  procedure 
should provide the symmetric stress-energy tensor 
density, $T^{\mu\nu}_{_N}(x)_{x\neq \xi}$,
 of the elementary particle-field matter.
Note that symmetric components of $g_{\mu \nu }$
are not independent  one from another,
$\delta g_{\mu \nu }$ = $\delta g_{\nu \mu }$, 
and we define
$\delta S  \equiv - \int dx^4{\sqrt {-g}} (T^{\mu\nu} 
\delta g_{\mu\nu}+ T^{\nu\mu} \delta g_{\nu\mu})/2  
 - \delta \int dx^4 {\sqrt {-g}} r/16\pi G.
$ When $i^\mu = const$, than
 $\delta ({ i}^\mu  V_{\mu }) $=${ i}^\mu\delta
[ g_{\mu\nu}dx^\nu (g_{\rho\lambda}dx^\rho dx^\lambda)^{-1/2} ]$
= $ [(\delta g_{\mu\nu}) { i}^\mu  dx^\nu/2ds] $
+ $ [(\delta g_{\nu\mu}) { i}^\nu  dx^\mu/2ds] $
= $({ i}^\mu dx^\nu  + { i}^\nu dx^\mu )
\delta g_{\mu \nu }/2ds$. The
 term $-{\sqrt {-g}}
( f^{\mu \nu }_{_N}\delta W_{_N\mu\nu}+
 f^{\nu \mu }_{_N}\delta W_{_N\nu\mu}) / 16\pi$
 may be transformed  into  $ ({\sqrt{-g}}\nabla_\nu
 {f}_{_N}^{\nu\mu}
 /4\pi) \delta  V_{_N\mu }$ under the integral.
     The contravariant metric tensor is related to the covariant one,
{\it i.e.} $\delta g^{\alpha \beta }
= - g^{\alpha \mu }g^{\beta \nu}\delta
g_{\mu \nu }$ $-$ $g^{\alpha \nu }g^{\beta \mu }\delta g_{\nu \mu }   $
$= - \delta g_{\mu \nu }(g^{\alpha \mu }g^{\beta \nu } + g^{\alpha \nu  }
g^{\beta \mu })$,
 $\delta {\sqrt {-g}} =  {\sqrt {-g}}( g^{\mu\nu}
\delta g_{\mu\nu} + g^{\nu \mu }\delta g_{\nu \mu }) / 2$
= $ {\sqrt {-g}}( g^{\mu\nu}
+ g^{\nu \mu })\delta g_{\mu \nu } / 2  $, with ${\sqrt {-g}}
= {\sqrt {g_{oo}}}$.

Finally, after the variation of (24) at the vertex free points with respect
 to  ${ \delta} g^{_N}_{\mu\nu}$ (and  $ \delta
 g^{_N}_{\nu\mu}$), one can derive an  Einstein-type equation,
\begin {eqnarray}
{1\over 8\pi G}\left ( r_{_N}^{ik}(x) - {1\over 2} g_{_N}^{\mu\nu} r_{_N}(x)
 \right )=   T_{_N}^{\mu\nu}(x)
\equiv  
     {{P^\mu_{_N}(x)
 I_{_N}^{\nu }(x)
+  P^\nu_{_N}(x)  I_{_N}^\mu (x)}\over 2} 
\nonumber \\
+ { W^{_N}_{\rho\lambda}(x)
\over 16 \pi}
[ g_{_N}^{\mu\nu} f_{_N}^{\rho\lambda}(x) -
2g_{_N}^{\mu\rho}f_{_N}^{\nu\lambda}(x)
- 2g_{_N}^{\nu\rho}f_{_N}^{\mu\lambda}(x)],
\end {eqnarray}
with the stress-energy tensor density $T_{_N}^{\mu\nu}(x) \equiv
T_{_N}^{\mu\nu}(x)_{x\neq\xi_{_N},\xi_{_N}}$
of the microscopic particle states and their self fields at all vacuum
points $x\neq \xi_{_N},\xi_{_K}$.

A trace of the stress-energy tensor density of elementary particle-field
matter in (33) always vanishes in vacuum points, $g^{_N}_{\mu\nu}
T_{_N}^{\mu\nu} \equiv 0$. This means from (33) that
the scalar curvature of the proper four-space is also absent,
$r_{_N}(x) \equiv 0$.
 Recall that the Rainich - Misner criterion
for   unified theories [21,22] dismisses scalar curvature terms in the
initial dynamical equations. The Ricci metric curvature,
$R^{\mu\nu}(x)$, will be introduced below for the macroscopic Einstein
 equation
after averaging the microscopic states over the global ensemble of matter.

By taking into account the vector variational equation (25), one may rewrite
the tensor equation (33) in the following form,
\begin {equation}
 r^{ik}(x) = 
  (\partial_\rho V_\lambda - \partial_\lambda V_\rho)
[g_{_N}^{\mu\rho} (-G m_{_N}f_{_N}^{\nu\lambda})
+  g_{_N}^{\nu\rho}(-G m_{_N}f_{_N}^{\mu\lambda})
- 2^{-1}g_{_N}^{\mu\nu} (-G m_{_N}f_{_N}^{\rho\lambda})].  
\end {equation}
From here the traceless tensor curvature $r^{\mu\nu}_{_N}(x)$ arises 
only for microscopic states with finite vorticity density, $\partial_\rho
 V_\lambda (x)\neq \partial_\lambda V_\rho (x)$, that maintains the tensor
  rank 2 of
the Einstein equation.

The variations of (24) with respect to the field co-ordinate 
of elementary matter, $\delta x_{_N}^\mu$,  
lead to its equation of motion in 
all vacuum points $ x \neq \xi_{_N}, \xi_{_K}$,
\begin {eqnarray}
2\nabla^{_N}_\mu   g^{_N}_{\rho \nu} T_{_N}^{\mu\rho} = 
{{\nabla^{_N}_\rho f^{\rho \lambda}_{_N}}\over 4\pi }
W^{_N}_{\lambda\nu} +
{{\nabla^{_N}_\rho W^{\rho \lambda}_{_N}  }\over 4\pi} f^{_N}_{\lambda\nu}
 =  i^\lambda_{_N} W^{_N}_{\lambda\nu}
 \nonumber \\
= m_{_N}w^{_N}_{\mu\nu}(x)i_{_N}^\mu (x) + [m_{_N}G_{\mu\nu}(x)
 + q_{_N}F_{\mu\nu}(x)]{{\nabla^{_N}_\rho f^{\rho\mu}_{_N}(x)}\over 4\pi} = 0, 
\end {eqnarray}
where we used variational equations (25), (28), 
and equalities 
$W^{\rho\lambda}\nabla_\nu f_{\rho\lambda} \equiv
 2W^{\rho\lambda}\nabla_\rho f_{\nu\lambda}$, 
$f^{\rho\lambda}
\nabla_\nu W_{\rho\lambda} \equiv 2f^{\rho\lambda}\nabla_\rho W_{\nu\lambda}$,
$\nabla^{_N}_\mu  g^{_N}_{\rho \nu} r_{_N}^{\mu\rho} \equiv 2^{-1}\partial
r_{_N} = 0$. By contracting (35) with $i^\nu_{_N}$ one finds an equality,
$i^\nu_{_N}i^\lambda_{_N} W^{_N}_{\lambda\nu}=0$, {\it i.e.} there are only
 three independent equations in this four-component
conservation law. 

In fact, (35) is a geodesic equation for microscopic particle states 
in vacuum points $x \neq \xi_{_N},\xi_{_K}$.
By integrating  $i^\lambda_{_N}(x) W^{_N}_{\lambda\nu}(x)=0$ over 
all space points, one can derive the macroscopic geodesic equation (22) 
for the averaged, localized particle state. A self-force $f^{_N}_\nu (x)$ $=$
 $ (q^2_{_N}$ $-$ $Gm^2_{_N}) f^{_N}_{\nu\mu}(x)\nabla^{_N}_\rho
 f_{_N}^{\rho\mu}(x)/4\pi $
$= -\nabla^{_N}_\mu \Theta^\mu_{\nu _N}(x)$, with $\Theta^\mu_{\nu _N}(x)
\equiv
(q^2_{_N}-Gm^2_{_N})[4f^{\mu\rho}_{_N}(x)
f_{\rho\nu}^{_N}(x)$ + $
\delta^\mu_\nu f^{\rho\lambda}_{_N}(x)f_{\rho\lambda}^{_N}(x)]$,
in the equation (35) is related locally to the plain vorticity
$w^{_N}_{\nu\mu}$ of the microscopic particle state in vacuum. This vacuum
 self-action on the extended
particle states is relevant to dynamics of the averaged, localized particle
 in the macroscopic equation (22). The total
space integral of the vacuum self-force $f_{_N}^\mu(x)$ is responsible for
the observed radiation damping.

In closing we may conclude from (26) and (33) that the microscopic mass and
 charge four-currents and the trace of the microscopic stress-energy tensor
  density of every elementary particle and its self fields are balances
  strictly to zero at all vacuum points. The point vertex $\xi_{_N}$ is
   the only unscreened mass/charge/energy carrier, which can directly be
    observed in practice.

\bigskip \bigskip
\noindent {\bf 6. The global superposition of microscopic states}

\bigskip 
 Even though the covariant equations are four-dimensional in the proper
four-space $x^\mu_{_N}$, dynamics
of matter depends on the development parameter, and there must be a
three-dimensional
picture as seen by an observer. This motivates us to pick out 
the coordinate world time $x^o=x^o_{_N}$ and the absolute time interval
$dt = |dx^o| > 0$ for the evolution of matter at all points of the world
 three-space ${\bf x}$.

Recall that we operate in (25) with the microscopic four-flow density
$i^\mu_{_N}(x)$ of virtual particle
fluctuations at all vacuum points $x\neq\xi_{_N}, \xi_{_K}$,
\begin {eqnarray}
\int\!{{dx^\nu}\over dp}{{\delta^4(x-\xi_{_N}[p])}\over 
{\sqrt {-g}}} dp \equiv  \int\!{{dx^\nu}\over d\xi_{_N}^o[p]} 
{{\delta^3 ({\bf x}-{\mbox {\boldmath $\xi $}}_{_N}[p])}
 \over {\sqrt {\gamma}}}{{\delta (x^o-\xi_{_N}^o[p])}
\over{\sqrt{g^{_N}_{oo}}}}d\xi_{_N}^o[p]
\nonumber \\ \equiv i^\nu_{_N}(x)  
= {\cases {
 i^i_{_N}(x) =  \delta^3
({\bf x}-{\mbox {\boldmath $\xi $}}_{_N}[x^o] ) 
 {{d x^i } / { \sqrt {g^{_N}_{oo}(x)} } dx^o }  \cr
i^o_{_N}(x) = 
{ \delta}^3
({\bf x}-{\mbox {\boldmath $\xi $}}_{_N} [x^o]) 
 / {  {\sqrt {g^{_N}_{oo}(x) }}  }\cr
}}.
\end {eqnarray}   

This four-flow density of the extended particle is locally bound with
 its
microscopic fields. A formal transition from the virtual, microscopic
 states to the point particle and its retarded/advanced macroscopic fields,
  available for measurements,
changes the structure of the particle flow density, $ i^i_{_N}(\xi) = 
{ \delta}^3
({\bf x}-{\mbox {\boldmath $\xi $}}_{_N} [x^o] ) 
 {{ d  \xi^i }/ {\sqrt {g^{_N}_{oo}(\xi) }}}  dx^o $ 
and $ i^o_{_N}(\xi) = 
{ \delta}^3
({\bf x}-{\mbox {\boldmath $\xi $}_{_N} [x^o]})$ 
 $  / {\sqrt {g^{_N}_{oo}(\xi) } }   $ in the Maxwell-type equations
 (25)-(25a).   Then the macroscopic electromagnetic and gravitational
  self fields exhibit retarded, $s_{_N}(x, \xi_{_N}[\tau_{+}]) = 0$, and
   advanced, $s_{_N}(x, \xi_{_N}[\tau_{-}]) = 0$, zero four-interval
   behaviour with respect to the point particle at $\xi_{_N}$.
Formally one may say  that these fields with the opposite evolution
directions are induced by outgoing ($dx^o > 0$)  and incoming  ($dx^o < 0$)
 virtual particle
fluctuations.   
There are no local intersections of point classical particles,
$\xi_{_N}\neq\xi_{_K}$. But there is a global intersection of their
 extended microscopic states.  Therefore one may derive  microscopic
 dynamic equations with local intersections of the extended particle
  states and read these
equations  in terms of averaged, macroscopic functions for point particles
 and retarded/advanced fields.

Every considered world space point  ${\bf x}$   can be related  to vertices 
 of different nonlocal particles by zero-interval conditions.
In other words, extended states of different particles can cross one common
world space point
 ${\bf x} \equiv {\bf x}_{_1},{\bf x}_{_2}..., {\bf x}_{_N}, ...$
like light of distant stars crosses the Earth at any fixed time.
Intersection of microscopic states of all nonlocal particle-field objects
 in one common (world)
3D space ${\bf x}$ may be described under one common 
co-ordinate  time $x^o(\bf x)$, which is independent from proper parameters
 of different objects.
All proper three-spaces $x_{_K}^i$, 
associated with different particles K, have 
universal 3D metrics with  $\gamma^{_K}_{ij} = \delta_{ij}$, 
due to (15).
For this reason 
the flat coordinate space ${\bf x}$ with the universal  3D interval $dl
={\sqrt {\delta_{ij}dx^id x^j }} > 0$ and the flat co-ordinate time with the
 absolute  1D interval $dt =|dx^o({\bf x})| = {\sqrt {\delta_{oo}dx^o d x^o }}
  > 0$, are quite appropriate to apply to all matter. Recall that
the four-interval $|ds_{_N}(x)| = {\sqrt {g^{_N}_{\mu\nu}dx^\mu dx^\nu }} > 0$
depends on local fields and, consequently,
on the path parameters $p_{_N}$ of elementary particles. 

One may say that the world space ${\bf x}$ is a flat 3D projection
of the infinite kaleidoscope of curved nonlocal 4D particle-field objects,
while $|dx^o({\bf x})|$ is simply an absolute world rate for stage set
changes on this invisible 3D screen.
Every vacuum point of the world space  $ {\bf x} $
at any particular  moment $x^o({\bf x})$ may contain  encountered
intersections of  microscopic particle-field states with zero
 mass/charge/energy densities. 
We write this statement by summing the microscopic equations (26) and (33)
over the global intersection of all elementary objects.
This leads to the following microscopic equations and their macroscopic
 generalizations for the total mass four-current density (respectively
 for virtual, nonlocal particles and point, classical particles associated
with a selected vacuum point $x \neq \xi_{_N},\xi_{_K}, ...$), 
\begin {equation}
  \sum^{all}_{_N}m_{_N} {\sqrt {g^{_N}_{oo}}}i_{_N}^\nu
(x)_{x\neq\xi_{_N}} = 
  -{{1}\over 4\pi G}\sum^{all}_{_N}
\partial_\mu {{(-Gm_{_N}){\sqrt {g^{_N}_{oo}}}f_{_N}^{\mu\nu}(x)_{x\neq
\xi_{_N}} }  }
 \end {equation}
$$\mu_o({\bf x})
{{d{x}^\nu}\over {\sqrt {{\tilde g}_{oo}}}dx^o} =
 -{1\over 4\pi G{\sqrt {{\tilde g}_{oo}}}}{\partial}_\mu {\sqrt {{\tilde
  g}_{oo}}} G^{\mu\nu}(x), \eqno (37a)
$$
for the  total electric four-current density, 
\begin {equation}
 \sum^{all}_{_N}q_{_N}{\sqrt {g^{_N}_{oo}}}i_{_N}^\nu
(x)_{x\neq\xi_{_N}} = 
{1\over 4\pi }  
  \sum^{all}_{_N}
\partial_\mu q_{_N}{\sqrt {g^{_N}_{oo}}}f_{_N}^{\mu\nu}(x)_{x\neq \xi_{_N}}, 
\end {equation}
$$\rho_o({\bf x})
{{dx^\nu}\over {\sqrt {{\tilde g}_{oo}}}dx^o} = {1\over 4\pi {\sqrt {{\tilde
 g}_{oo}}}}
{\partial}_\mu {\sqrt {{\tilde g}_{oo}}} F_{_N}^{\mu\nu}(x), \eqno (38a)$$
and for the total stress-energy tensor density,
\begin {eqnarray}
\sum^{all}_{_N}{{m_{_N}{\sqrt { g^{_N}_{oo}}} }\over 2}
 \left( { i}_{_N}^\mu(x) V^\nu
+ {i}_{_N}^{\nu }(x)
 V^\mu  \right)_{x_{_N}\neq
 \xi_{_N}}
 \nonumber \\
\equiv \sum^{all}_{_N}   \left \{  {{   { \delta}^3
 ({\bf x}-{\mbox {\boldmath $\xi $}  }_{_N})}}m_{_N}
{{ dx^\mu}\over  dx^o} {{ dx^\nu}\over  ds_{_N}}  \right \}_{{\bf x}\neq
 {\mbox {\boldmath $\xi $}  }_{_N}}
 = {1\over 8\pi G}\sum^{all}_{_N}{\sqrt { g^{_N}_{oo}}}
r^{\mu\nu}_{_N}(x)_{x_{_N}\neq \xi_{_N}} 
   \nonumber \\
-    {1\over 8\pi G}\!\sum^{all}_{_N}[
 -G m_{_N}{{d x^\mu}\over  d{ s_{_N}}}
  {\sqrt { g^{_N}_{oo}}}\nabla^{_N}_\rho  {f}_{_N}^{\rho \nu}(x)
- Gm_{_N}{{ d x^\nu}\over 
 d{ s_{_N}}}{\sqrt { g^{_N}_{oo}}}\nabla^{_N}_\rho  {f}_{_N}^{\rho \mu}(x) 
\nonumber \\
-  Gm_{_N}M^{_N}_{\rho\lambda}(x){\sqrt { g^{_N}_{oo}}}[ g_{_N}^{\mu\rho}
f_{_N}^{\nu\lambda}
(x) + g_{_N}^{\nu\rho}f_{_N}^{\mu\lambda}
(x) - {{g_{_N}^{\mu\nu}}\over 2}f_{_N}^{\rho
\lambda}(x)] ]_{x\neq
 \xi_{_N}}
  \nonumber \\
+  {{F_{\rho\lambda}(x)}\over 8\pi}\!\sum^{all}_{_N}\{q_{_N}{\sqrt
 { g^{_N}_{oo}}}\![ g_{_N}^{\mu\rho}f_{_N}^{\nu\lambda}
(x) + g_{_N}^{\nu\rho}f_{_N}^{\mu\lambda}
(x) - {{g_{_N}^{\mu\nu}}\over 2}f_{_N}^{\rho
\lambda}(x)] \}_{x\neq
 \xi_{_N}},
\end {eqnarray}
$$ \mu_o({\bf x})
 {{ dx^\mu}\over {\sqrt {{\tilde g}_{oo} } } dx^o } {{ d{x}^\nu}\over
  d{\tilde s} } =
 {1\over 8\pi G} \left (
  R^{\mu\nu}(x) - {{\tilde g}^{\mu\nu}\over 2} {\tilde g}_{\rho\lambda} 
R^{\rho\lambda}(x)
   \right )$$
$$ - {1\over 16\pi}[  2{\tilde g}^{\mu\rho} F_{\rho\lambda}(x) {\tilde
 F}^{\lambda\nu}(x) +
 2{\tilde g}^{\nu\rho} F_{\rho\lambda}(x) {\tilde F}^{\lambda\mu}(x)
 + {\tilde g}^{\mu\nu}F_{\rho\lambda}(x){\tilde F}^{\rho\lambda}(x) ].
  \eqno (39a)
$$

Here we introduced  global compositions of elementary macroscopic
(advan\-ced/retarded) fields,
 $G^{\mu\nu}(x)$ $\equiv$ $ [{{ {\tilde g}_{oo}}}(x)]^{-1/2}\sum^{all}_{_N}
 (-Gm_{_N}) {\sqrt {  g^{_N}_{oo}}}f_{_N}^{\mu\nu}(x)^{s_{_N}
[\tau_{-} ]={_0}}_{x\neq \xi_{_N}[\tau_{-}]}$,
$F_{\rho\lambda}(x)\!\equiv\!\sum^{all}_{_N}\!q_{_N}f^{_N}_{\rho\lambda}
(x)^{s_{_N}[\tau_{+} ]={_0}}_{x\neq \xi_{_N}[\tau_{+}]}$,
${\sqrt{{\tilde g}_{oo}(x)}}\!F^{\mu\nu}(x)\!\equiv\!\sum^{all}_{_N}
{\sqrt { g^{_N}_{oo}}}q_{_N}f_{_N}^{\mu\nu}(x)^{s_{_N}
[\tau_{+} ]={_0}}_{x\neq \xi_{_N}[\tau_{+}]}$, and 
$ {\sqrt{{\tilde g}_{oo}(x)}}{\tilde g}^{\mu\nu}(x){\tilde F}^{\rho\lambda}
(x)$ $\equiv$
$ \sum^{all}_{_N}  {\sqrt {  g^{_N}_{oo}}} g_{_N}^{\mu\nu}(x)
q_{_N}f_{_N}^{\rho\lambda}(x)^{s_{_N}
[\tau_{+} ]={_0}}_{x\neq \xi_{_N}[\tau_{+}]}$ 
 in order to use them in the
 macroscopic equations at vacuum points $x \neq \xi_{_N}$. Notice
that all elementary microscopic fields contribute to these macroscopic fields.
This is not a case for the macroscopic particle densities $\mu_{o}(x)$ 
and $\rho_o(x)$ in vacuum (field) points, because the macroscopic particle
 state (resulting from averaging over the microscopic states under small
 macroscopic time scales) is localized at one point $\xi_{_N}$. The global
  sums of the microscopic densities $ \sum^{all}_{_N}
   m_{_N} {\delta}^3
 ({\bf x}-{\mbox {\boldmath $\xi $}}_{_N})$ and 
  $\sum^{all}_{_N}  q_{_N} {\delta}^3({\bf x}-
{\mbox {\boldmath $\xi $}}_{_N})$
at the field point ${\bf x}$ are to be replaced under macroscopic time  
with a summing over those particles, which are localized in a small
macroscopic  volume $\Omega \rightarrow 0 $ around  ${\bf x}$,  
  $ \sum^{all}_{_N}
   m_{_N} {\delta}^3
 ({\bf x}-{\mbox {\boldmath $\xi $}}_{_N})$ $\rightarrow$ ${ \mu}_o({\bf x})$
 $ \equiv$
$ \sum^{\Omega}_{_N} m_{_N} ({\mbox {\boldmath $\xi $}}_{_N}) / \Omega $,
and $ \sum^{all}_{_N}
  q_{_N} {\delta}^3
 ({\bf x}-{\mbox {\boldmath $\xi $}}_{_N})$ $\rightarrow$
${ \rho}_o({\bf x})$ $ \equiv$  $\sum^{\Omega}_{_N} 
   q_{_N} ({ \mbox {\boldmath $\xi $} }_{_N}) / \Omega $.

The macroscopic metric tensor ${\tilde g}_{\mu\nu}(x)$ for an elementary 
ensemble of all point particles in $\Omega$ is defined in the field, vacuum
 point $x\neq\xi_{_N}$  by (15) with an elementary  mass
 $m_{_N} = {\mu}_o({\bf x})\Omega$ and an elementary charge  $q_{_N} =
{\rho}_o({\bf x})\Omega$.
Space is flat for macroscopic gravitation and electrodynamics, because
${\tilde g}_{oi}{\tilde g}_{oj} {\tilde g}^{-1}_{oo} - {\tilde g}_{ij}=
 \delta_{ij}$,
${\tilde \gamma} = 1$, and  
${\sqrt {-{\tilde g} }} $ 
$\equiv$ $ {\sqrt { {\tilde \gamma} {\tilde g}_{oo}} }$
 = ${\sqrt {{\tilde g}_{oo}}}$.

It is worth to repeat that all point sources-outlets are excluded from the
 original microscopic  equations (37), (38), and (39) for vacuum points.
 Contrary to the continuous density of the microscopic, extended particle
  states,
the density of  the macroscopic,  point  particles
 in one selected field point $\bf x$ is meaningless.
One should not neglect this obvious fact under interpretation 
of the field equations for electrodynamics or gravitation.

The macroscopic Ricci tensor ${R}^{\mu\nu}(x)$ 
$\equiv  {\cal G}^{\mu\nu}(x) - 2^{-1}{\tilde g}^{\mu\nu}
{\tilde g}_{\rho\lambda}{\cal G}^{\rho\lambda}(x)$ in (39a)
and  the Einstein tensor 
${\cal G}^{\mu\nu}(x)$ are
defined  by a following superposition of the elementary traceless
tensors $r^{\mu\nu}_{_N}$ and advanced gravitational fields,
\begin {eqnarray}
  {\cal G}^{\mu\nu}({x}) \equiv  
{{1} \over {\sqrt {  {\tilde g}_{oo} }}  }\!\sum^{all}_{_N} 
{\sqrt {g^{_N}_{oo} } }\left ( r^{\mu\nu}_{_N}
+ Gm_{_N} [ {{d x^\mu}\over  d s_{_N}}
  \nabla^{_N}_\rho  {f}_{_N}^{\rho \nu}(x)
+ {{d x^\nu}\over  d{s_{_N}}}\nabla^{_N}_\rho  {f}_{_N}^{\rho \mu}(x)]\right.
\nonumber \\
+ Gm_{_N}M^{_N}_{\rho\lambda}(x)[ g_{_N}^{\mu\rho}{f}_{_N}^{\nu\lambda}(x)
 + g_{_N}^{\nu\rho}{f}_{_N}^{\mu\lambda}(x)
 - {{g_{_N}^{\mu\nu}}\over 2}{f}_{_N}^{\rho
\lambda}(x)] \Biggr )^{s_{_N}[\tau_{-}]={_0}}_{x\neq \xi_{_N}[\tau_{-}]}.
\end {eqnarray}

The scalar Ricci "curvature",  $ {\tilde g}_{\rho\lambda}{R}^{\rho\lambda}(x)
 \equiv R(x) = - {\cal G} (x)\equiv - {\tilde g}_{\rho\lambda}{\cal
  G}^{\rho\lambda}(x)$, originates from advanced, incoming elementary
   fields. Recall that these gravitational fields were induced in order
    to screen microscopic mass four-current densities of nonlocal particles
    in vacuum points, rather than to curve flat three-space. One may say the
     scalar "curvature"
$R$ of macroscopic fields balances the mass
 density $\mu_o ({\bf x})$ of macroscopic particles in vacuum points. 
This "curvature" may be found from
 the macroscopic gravitational equation (39a) or directly from the original
 microscopic
 equality $\sum_{_N} {\sqrt { g_{oo}^{_N} }}g^{_N}_{\mu\nu} [{8\pi G}
 T^{\mu\nu}_{_N}(x) - r^{\mu\nu}_{_N}(x)]_{x\neq \xi_{_N},\xi_{_K}}\equiv
  0 $,
\begin {equation}
 R(x) \!=\!
{ {-8\pi G  \mu_o({\bf x}) d {\tilde s}}\over 
{\sqrt {{\tilde g}_{oo}} }  d x^o }
=\!{8\pi\over {\sqrt {{\tilde g }_{oo}}  }} 
\sum^{all}_{_N}(-G m_{_N})\! 
 {  {g^{_N}_{\mu\nu}dx^\mu dx^\nu  }\over ds_{_N} dx^o}
   \delta_{_N}^3 ({\bf x}- {\mbox {\boldmath $\xi$}}_{_N}).
\end {equation}

One may find from (14) a global sum for the four-momentum densities,
${\cal P}^{mic}_\mu (x)\equiv \sum_{_N}^{all} {\delta}^3
 ({\bf x}-{\mbox {\boldmath $\xi $}  }_{_N}) m_{_N}g^{_N}_{\mu\nu}dx^\nu
/ ds_{_N}$, of all microscopic states in any vacuum point and a 
macroscopic four-momentum density, ${\cal P}^{mac}_\mu (x)$
 $\equiv \mu_o (\bf x)$
 $ {\tilde g}_{\mu\nu} dx^\nu / d{\tilde s} $, of point particles
in the nearest vicinity of the selected vacuum 
point  $ x=\{{\bf x}; x^o\} \neq \xi_{_N}, \xi_{_K}$, 
\begin {eqnarray}
 {\cal P}^{mic}_\mu (x) = \sum_{_N}^{all} m_{_N}{\cal V}_{_N\mu}(x)
 {\delta}^3
 ({\bf x}-{\mbox {\boldmath $\xi $}  }_{_N})
  +  B_\mu (x)\sum_{_N}^{all} m_{_N}{\delta}^3
 ({\bf x}-{\mbox {\boldmath $\xi $}  }_{_N})
\nonumber \\ +
A_\mu (x) \sum_{_N}^{all} q_{_N}{\delta}^3
 ({\bf x}-{\mbox {\boldmath $\xi $}  }_{_N}) + 
\sum_{_N}^{all} m_{_N}{\delta}^3 ({\bf x}-{\mbox {\boldmath $\xi $}}_{_N})
\partial_\mu \phi_{_N}, 
\end {eqnarray}
$${\cal P}^{mac}_\mu (x) = 
\mu_o ({\bf x}) [ {\tilde { \cal V}}_\mu + B_\mu (x)] + \rho_o
({\bf x})A_\mu (x)  + \mu_o ({\bf x}) \partial_\mu {\tilde \phi}_{_N}(x).
 \eqno (42a) $$

The local field potentials $A_\mu (x)$ and $B_\mu (x)$ contribute with the
 identical signs to the macroscopic four-momentum density (42a), but with
 the opposite signs to (37a) and (38a) for the mass and charge four-current
  densities, respectively. By considering these macroscopic relations one
 finds that the electric current densities can screen locally external
 electromagnetic fields, while the mass current densities can be responsible
  only for oscillating solutions of external gravitational fields without
  their damping. In other words there are
no screens in practice for macroscopic gravitational fields.
 The physical reason may be traced as follows: the point charge can interact
  with casual electromagnetic waves, while advanced disturbances of the
  incoming gravitational field never arrive from infinity in finite times
  and "inform" the point mass in the laboratory.

The requirement for finite magnitudes for all material
densities supports our introduction of the extended particle states in 
vacuum. It seems very unlikely that it is possible to overcome
the problem of divergence in electrodynamics, for example, by applying
 the classical paradigm of the point charge to microscopic field theory.
  The developed microscopic approach to gravitation is quite consistent
  with
the cited Einstein's statements, which worth to be repeated: 
"A coherent field theory requires that all its elements
be continuous ... And from this requirement arises the fact that the
material particle
has no place as a basic concept in a field theory. Thus, even apart
from the fact
that it does not include gravitation, Maxwell's theory cannot be considered 
as a complete
theory" (translation [19]). Our dual formalism for the
macroscopic/microscopic particle within the extended elementary
object is a predicted way to introduce the "continuous element" into the
 classical field equations for electrodynamics and gravitation.

\bigskip\bigskip

\noindent {\bf 7. Conclusion }

\bigskip

The particle mass four-current $m_{_K}i^\mu_{_K}(x)$ is the origin of the
 incoming gravitational field $(-Gm_{_K}f^{_K-}_{\mu\nu})$ and the vector
  gravitational force $m_{_N}(-Gm_{_K})V_{_N}^\nu f^{_K-}_{\mu\nu}$.
According to General Relativity the electromagnetic and gravitational  parts
 of interactions cannot compensate each other in the state of general motion
  due to the different tensor nature of these interactions.
The electromagnetic and gravitational external potentials for the charged
 particle have a unified retarded/advanced structure,
$\sum_{_K}^{_K\neq _N} (q_{_N}q_{_K}a^+_{_K\mu} -
 G m_{_N}m_{_K}a^{-}_{_K\mu})$,
in our approach, which lead the vector balance of electromagnetic and
gravitational "instantaneous" forces for a two-body system
with $q_{_1}q_{_2} = Gm_{_1}m_{_2}$.  

The absence of aberrations of the "instantaneous" Newton and Coulomb forces,
based on the advanced and retarded field potentials, respectively, does
 not contradict to the finite speed c of gravitational and electromagnetic
  waves.
Could Sun's incoming gravitational wave come to the laboratory
from infinity, it would have an advanced aberration 20 arc seconds with
 respect to the attraction force direction and 40 arc seconds with respect
to the retarded aberration of Sun's light.

The developed coupling of electromagnetic and gravitational vector
interactions and the integration of the particle into its field structure
 satisfy   the predicted double unified criterion [19], as well as  the
 Rainich-Misner known criterion [21,22] for the unified field theory. All
  point sources-outlets  are excluded from the continuous particle-field
  equations in  agreement with Einstein's approach [23,19,20] to the
   continuum theory, and all  physical magnitudes in vector electrogravity
    are free from divergences.

We accepted only  retarded potentials for outgoing electromagnetic fields
with respect to their point sources and only advanced potentials for
incoming gravitational fields with respect to their point outlets.  The
opposite directions of outgoing and incoming spherical fields explain
the repulsion of identical charges and the attraction of masses in the
 vector field theory.
The unified vector nature of electrogravity corresponds to the unified
 spin-1 photon/graviton approach to electromagnetic/gravitational waves.
 Our theory predicts the absence of metric modulations of flat world space
in  disagreement with the tensor (spin-2) gravitational wave
propounded by General Relativity.

The electromagnetic/gravitation waves have zero charge/mass four-current
 densities. These pure field objects do not couple to each other,
contrary to particles with finite four-currents.  
The vector gravitational wave, {\it i.e.} the electromagnetic antiwave in
 our approach, may be formally associated with a negative-energy hole in
 the
retarded radiation spectrum due to the inverse
  time rate for the advanced, incoming field matter. 
The electromagnetic repulsion of identical charges is  mediated by the
 virtual particles (photons) in quantum field theory, where the gravitational
  attraction of masses may be formally mediated by holes (gravitons) in the
  retarded field spectrum.  This approach creates a clear perspective  to
  quantize the gravitational spin-1 field in a full analogy with the
  successfully quantized electromagnetic field.

Based on electrodynamics and its references, vector gravitation becomes
 a self-contained theory, which may be applied to practice without
 references on other gravitomechanical theories. Nevertheless, the
 developed approach to gravitation coincides with the Newton theory in
 the non-relativistic limit and with the concept of flat world space $x^i$
  and the absolute time rate $dt(x^i)$ = $dt(x^j)$ in all space points. The
   linear relativistic
corrections to Newton's motion in the present theory with flat three-space
coincide quantitatively with the similar corrections of General Relativity,
 but our nonlinear solutions for strong fields do not lead to
GR's black holes.
 The available observations of all known kinds of interactions and all
 conservation laws do not contradict to Euclidean geometry of the world
  space, which we considered as a material continuum
 of the microscopic particle states and their fields with vanishing
 charge/mass/energy densities at all space points except at point
 sources-outlets.

 The main relativity tests (the gravitation light bending by the Sun,
  the radar echo delay, and the Mercury perihelion precession) can be
   quantitatively explained under flat three-space with the absolute time
 rate in all space points.  The superfluid microscopic states of the
 nonlocal particle in
vacuum are consistent with the Aharonov-Bohm phenomenon [24].
The nearly isotropic microwave background radiation corresponds to
the spatial flatness of the Universe [3] that is in an agreement with
the metric tensor  (15) for elementary matter.

One may ask what is the antimass in the present theory with incoming
gravitational fields. A spontaneous split of the zero-mass "object"
should create at least a couple of mass-antimass outlets with their
 advanced fields, which moves separately toward these outlets.
The electric charge source and its antisource  attract each other
due to their outgoing fields and tend to collapse, while the mass outlet
 and anti-outlet  repulse each other due to their incoming fields and tend
  to disintegrate. The repulsion of galaxies with the mass and the antimass
   should lead to an increase of their mutual velocities and to  the
   Universe expansion with acceleration.

The dual approach to the microscopic/macroscopic particle and its fields
in (25)-(25a), for example, is based on their instantaneous
 (C/c=$\infty$) virtual
fluctuations with zero energy tensor density within the nonlocal
particle-field object. This is
quite consistent with instantaneous reshaping of Newton/Coulomb forces
 and
generation of advanced macroscopic waves in infinity. These infinitely
 distant gravitational waves have the light speed c, like the retarded
  electromagnetic waves, and will never reach the laboratory in finite
   time intervals. Therefore the advanced gravitation wave do not violate
    the casual requirements of Special Relativity for macroscopic fields
    and particles. Nonetheless, such generation-at-a-distance of
     gravitation waves "there"
     (in infinity) requires energy consumption by accelerated masses "here"
      and leads, for example, to radiation damping for rotating
       binary pulsars.
More general, the unified approach to elementary matter through  the nonlocal
particle-field object N, which is self-enclosed into infinite four-space
 $x^\mu_{_N}$,  satisfies to the Mach's principle - matter there
  (gravitational waves)
   governs inertia (of point particles) here  [25, 8].  Anyway, the
        nonlocal nature of the Universe, confirmed last century in different
   tunnel and light experiments,  for example [26, 27], is open for
    further exploration and discussions.

\bigskip 
\bigskip
\noindent {\bf References}

\bigskip
\noindent [1]  Einstein A 1916 {\it Annalen der Physik} ${\bf 49}$ 769

\noindent [2] Schwarzschild K 1916 
{\it Kl. Math.-Phys. Tech.} (Sitzungber. Deut. Akad. Wiss., Berlin) 424,

\noindent [3] De Bernardis P {\it et al.} 2000 {\it Nature} ${\bf 404}$
955, Lange A {\it et al.} 2001 {\it Phys. Rev.} ${\bf D63}$ 042001,
 Hanany S {\it et al.} 2000  {\it Astrophys. J.} ${\bf 545}$ L5  

\noindent [4]. Rosen N 1940  {\it Phys. Rev.} ${\bf 57}$ 147,
  Logunov A A 1995 {\it Physics - Uspekhi} ${\bf 38}$ 179

\noindent[5]  Landau L D  and  Lifshitz E M 1971
{\it The Classical Theory of Fields} (Oxford: Pergamon)

\noindent [6]  Weinberg S 1972 {\it Gravitation and Cosmology} (New York:
John Wiley and Sons )

\noindent  [7]  Wald R M 1984  {\it General Relativity} (Chicago: The
University of Chicago Press)

\noindent [8] Misner C W, Thorne K S and  Wheeler J A 1973
{\it Gravitation} (Freeman, San Francisco: Freeman)

\noindent [9] Will C M  
{\it Theory and Experiment in Gravitational Physics} 1981 (Cambridge:
Cambridge University Press)  

\noindent [10] Synge J L 1960  {\it Relativity: the General Theory}
(Amsterdam: North-Holland Publishing Company)

\noindent [11] Einstein A 1907 {\it Jahrb. f. Radioaktivit${\ddot a}$t
u. Elektronik} {$\bf 4$} 411

\noindent [12] Einstein A 1911 {\it Annal. der Phys.} (Leipzig) {\bf 35} 898 

\noindent [13] Pound R V and Rebka G A 1960 {\it Phys. Rev. Lett.} {\bf 4}
 337

\noindent [14]  Shapiro I I 1964 {\it Phys. Rev. Lett.} {\bf 13} 789 

\noindent [15] Williams J G {\it et. al.} {\it Phys. Rev. Lett.} {\bf 36}
 551

\noindent [16] Shapiro I I,   Counselman C C and  King R W 1976 {\it Phys.
 Rev. Lett.} {\bf 36} 555

\noindent [17] Hilbert D 1917 {\it Nachrichten von der Gesellschaft
der Wissenschaften zu Gottingen} {\bf 4}  21

\noindent [18] De Broglie L 1923 {\it Compt. Rend. Acad. Sci. Paris}
{\bf 174}, 506, 548, 630;  de Broglie L 1973 {\it Wave Mechanics: the First
 Fifty Years } eds W C Price, S S Chissick and T Ravensdale  (London:
Univer. of London King's College)

\noindent [19]  Tonnelat M - A 1966  {\it The Principles of Electromagnetic
 Theory
and Relativity} (Dordrecht:  Reidel Publishing Co)

\noindent [20] De Broglie L 1962 {\it New Perspectives in Physics}
(New York: Basic Books)

\noindent [21]  Rainich G Y 1925  {\it Trans. Am. Math. Soc.} {\bf 27}
106  

\noindent [22]  Misner C  and Wheeler J 1957
{\it Ann. of Physics} {\bf 2} 525  

\noindent [23] Einstein A 1956 {\it The Meaning of Relativity}
(Princeton: Princeton University Press) { Appendix {\uppercase\expandafter
{\romannumeral 2} }}

\noindent [24]  Aharonov Y and  Bohm D 1959  {\it  Phys. Rev.} {\bf 115}
 485 

\noindent [25] Mach E 1904  {\it Die Mechanik in ihrer Entwickelung
historisch-kritisch
darges\-tellt} (Leipzig: Brockhaus F A) p 236 

\noindent [26]. Aspect A,  Grangier P and  Roger G 1981 {\it Phys. Rev.
Lett. }{\bf 47} 460

\noindent [27].  Tittel W, Brendel J,  Zbinden H and Gisin N 1998 
{\it Phys. Rev. Lett.}  {\bf 81} 3563 

\end {document}